\documentclass[prd, aps, nofootinbib, preprintnumbers, showpacs, superscriptaddress,twocolumn]{revtex4}
\usepackage{latexsym}
\usepackage{amssymb}
\usepackage{amsfonts}
\usepackage{amsmath}
\usepackage[dvips]{graphicx}\usepackage{color}

\usepackage{ulem}

\renewcommand{\emph}[1]{{\it #1}}

\newcommand{\be}{\begin{equation}}  
\newcommand{\ee}{\end{equation}}
\newcommand{\bea}{\begin{eqnarray}}           
\newcommand{\eea}{\end{eqnarray}} 
\newcommand{\beqn}{\begin{eqnarray*}}
\newcommand{\eeqn}{\end{eqnarray*}}
\newcommand{\ba}{\begin{align}}
\newcommand{\ea}{\end{align}}
\newcommand{\cf}{\textit{cf.}}
\newcommand{\ie}{\textit{i.e.}~}
\newcommand{\eg}{\textit{e.g.}~}
\newcommand{\tg}{\tilde\gamma}
\newcommand{\tG}{\tilde\Gamma}
\newcommand{\tA}{\tilde A}

\newcommand{\dt}{(\partial_t - {\cal L}_\beta)\;}
\DeclareMathOperator{\tr}{\ensuremath{\mathrm{tr}}}

\def\half{\frac{1}{2}}
\def\de{\partial}
\def\lm{{\ell m}}
\def\o{{\rm o}}

\def\l{{\ell }}
\def\r{{\bar{r}}}

\def\DENSq{\mathbf{\varrho}}
\def\i{{\rm i}}
\def\I{{\cal I}}
\def\J{{\cal J}}
\def\U{{\cal U}}
\def\V{{\cal V}}
\def\M{{\cal M}}
\def\S{{\cal S}}
\def\X{{\cal X}}
\def\Y{{\cal Y}}
\def\W{{\cal W}}
\def\Z{{\cal Z}}
\def\e{{e}} 

\begin{document}


\title{Gravitational-Wave Extraction from Neutron-Star Oscillations:\\ 
  comparing linear and nonlinear techniques.} 

  \author{Luca \surname{Baiotti}}
  \affiliation{Graduate School of Arts and Sciences, 
    University of Tokyo, Komaba, Meguro-ku, 
    Tokyo, 153-8902, Japan}
  \affiliation{Max-Planck-Institut f\"ur Gravitationsphysik,
    Albert-Einstein-Institut, Potsdam-Golm,Germany}
  \author{Sebastiano \surname{Bernuzzi}}
  \affiliation{Dipartimento di Fisica, Universit\`a di Parma
    and INFN, Gruppo Collegato di Parma,
    via G.~B.~Usberti 7/A, 43100 Parma, Italy}
  \author{Giovanni \surname{Corvino}}
  \affiliation{Max-Planck-Institut f\"ur Gravitationsphysik,
    Albert-Einstein-Institut, Potsdam-Golm,Germany}
  \affiliation{Dipartimento di Fisica, Universit\`a di Parma
    and INFN, Gruppo Collegato di Parma,
    via G.~B.~Usberti 7/A, 43100 Parma, Italy}
  \author{Roberto \surname{De Pietri}}
  \affiliation{Dipartimento di Fisica, Universit\`a di Parma
    and INFN, Gruppo Collegato di Parma,
    via G.~B.~Usberti 7/A, 43100 Parma, Italy}
  \author{Alessandro \surname{Nagar}}
  \affiliation{Institut des Hautes Etudes Scientifiques, 91440 Bures-sur-Yvette, France}
  \affiliation{INFN, Sezione di Torino, Via Pietro Giuria 1, Torino, Italy}
  \affiliation{ICRANet, 65122 Pescara, Italy}
  
  \date{\today}
  
  \begin{abstract}
    The main aim of this study is the comparison of gravitational waveforms 
    obtained from numerical simulations which employ different numerical 
    evolution approaches and different wave-extraction techniques.  
    For this purpose, we evolve an oscillating, nonrotating,
    polytropic neutron-star model with two different 
    approaches: a full nonlinear relativistic simulation (in three dimensions)
    and a linear simulation based on perturbation theory.
    The extraction of the gravitational-wave signal is performed via three 
    methods: the gauge-invariant {\it curvature-perturbation} theory based on 
    the Newman-Penrose scalar $\psi_4$; the gauge-invariant 
    Regge-Wheeler-Zerilli-Moncrief {\it metric-perturbation} theory of 
    a Schwarzschild space-time; some generalization of the quadrupole 
    emission formula.
  \end{abstract}

  
  \maketitle
  
    
  \section{Introduction}
  \label{sec:intro}

  The computation of the gravitational-wave emission from
  compact sources like supernova explosions, neutron-star oscillations 
  and the inspiral and merger of two compact objects (like neutron stars or
  black holes) is one of the most lively subjects of current research in 
  gravitational-wave astrophysics. This goal may be pursued using different 
  numerical approaches. That is, (i) solving the {\it full set} 
  of coupled Einstein and matter equations; (ii) solving the {\it linearized}
  Einstein and matter equations around a fixed background, when such an approximation is valid.
  In the latter case, with the additional condition of spherical symmetry, 
  the formalism we employ is based on a multipolar 
  expansion and the computation of the gravitational waves directly follows
  from the knowledge of the perturbative metric multipoles $k_\lm$, $\chi_\lm$ 
  and $\psi_\lm$.
  On the other hand, extracting gravitational waveforms from a space-time
  computed numerically in a given coordinate system is a highly nontrivial 
  problem that has been addressed in various ways in the literature.
  In general, two routes have proven successful: (i) the gauge-invariant 
  {\it curvature-perturbation} theory based on the 
  Newman-Penrose~\cite{NP62} scalar $\psi_4$, and (ii) the 
  Regge and Wheeler~\cite{RW57}, Zerilli~\cite{Zerilli:1970se} 
  theory of {\it metric-perturbations} of a Schwarzschild 
  space-time, recast in a gauge-invariant framework 
  following the work of Moncrief~\cite{Moncrief:1974am}.

  The aim of our study is the computation of the gravitational waveforms emitted by
  the very controlled system constituted by a nonrotating polytropic 
  relativistic star that oscillates nonisotropically 
  around its spherically symmetric equilibrium configuration
  because of an axisymmetric perturbation. 
  Our aim is to follow two (complementary) calculation procedures. On one hand, we perform
  a full 3+1 numerical simulation of the system, \ie we compute a 
  numerical solution of the Einstein equations without approximations
  except those of the numerical method itself.
  Because of its generality, this approach allows us to analyze different 
  physical regimes, in particular,
  the case in which the ``perturbation'' is not small and nonlinear
  effects can play a relevant role with important consequences on the waveforms.
  On the other hand, we follow a perturbative approach based on the assumption 
  that the perturbation is ``small''. If this is the case, one can (i) expand 
  the metric around a fixed background (\ie the Tolman-Oppenheimer-Volkoff solution),
  (ii) retain only the linear term of this expansion and (iii) solve
  the {\it linearized} Einstein equations. In addition, since the star is 
  nonrotating, one can factorize the angular dependence by means of a 
  spherical-harmonic decomposition of the metric and matter fields, and, thus, only a 
  1+1 system of partial differential equations must be solved.
  
  The present work has much in common 
  with Refs.~\cite{Shibata:2003aw,Pazos:2006kz}, 
  where a comparison of different extraction techniques has been 
  performed. 
  Following the same inspiration of Ref.~\cite{Pazos:2006kz},
  we exploit perturbative computations to obtain ``exact''
  waveforms to compare with the numerical-relativity--generated
  ones. As done in Ref.~\cite{Shibata:2003aw},
  we use an oscillating neutron star as a test-bed system,
  but we consider a wider range of possible wave-extraction
  techniques.
  Since there is a copious literature dealing with 
  the problem of gravitational-wave extraction in
  numerical relativity, we prefer not to mention here the 
  main bibliographic references, but rather to address the 
  reader to the references in Refs.~\cite{Shibata:2003aw,Pazos:2006kz}
  and to the citations in the following text.  

  The article is organized as follows. In Sec.~\ref{sec:nsetp} we describe 
  the numerical time-evolution methods and the gravitational-wave 
  extraction techniques adopted.
  In Sec.~\ref{sbsc:init} we introduce our choice of initial data
  and Sec.~\ref{sec:res} is devoted to the
  presentation of our results. Conclusions that can be drawn from our
  results are discussed in Sec.~\ref{sec:concl}.
  
  Standard dimensionless units $c=G=M_\odot=1$ and a spacelike signature
  $(-,+,+,+)$ are used. Greek indices are taken to run from $0$ to $3$, 
  Latin indices from $1$ to $3$ and we
  adopt the standard convention for the summation over repeated indices.
  
  \section{The physical system and its numerical evolution}
  \label{sec:nsetp}

  In this section we present the main elements of the two evolutionary 
  approaches and discuss the three wave-extraction techniques mentioned
  in the introduction. 
  In our investigation we deal with the full set of Einstein equations 
  \begin{equation}
    \label{eq:Einstein}
    G_{\mu\nu} = 8 \pi T_{\mu\nu} \;,
  \end{equation}
  coupled to a perfect-fluid matter, with stress-energy tensor
  \begin{equation}
    \label{eq:stress-energy}
    T^{\mu\nu} = \rho \left(1 +\epsilon + \frac{p}{\rho}\right)  
    u^{\mu} u^{\nu} + p g^{\mu\nu},
  \end{equation}
  where $u^\mu$ is the fluid 4-velocity, $p$ is the fluid pressure,
  $\epsilon$ is the specific internal energy and $\rho$ is the rest-mass
  density, so that $\e = \rho (1+\epsilon)$ is the energy density in the
  rest frame of the fluid and $H=\rho(1+\epsilon)+p$ is the relativistic 
  specific enthalpy. The Einstein equations for the space-time must be
  supplemented by the relativistic hydrodynamics equations, namely, the conservation 
  law for the energy-momentum tensor $\nabla_{\mu} T^{\mu\nu}= 0$, the conservation 
  law for the baryon number $\nabla_{\mu} (\rho u^{\mu}) = 0$, and an
  equation of state (EOS) of the type $p=p(\rho,\epsilon)$. For the purpose
  of this work, we restrict our attention to the polytropic  
  (isoentropic) equation of state:
  \begin{equation}
    \begin{aligned}
      p        &= K \rho^\Gamma \, , \\ 
      \epsilon &= \frac{K}{\Gamma-1} \rho^{\Gamma-1} \, , 
    \end{aligned}
    \label{eq:EOS}
  \end{equation} 
  with parameters $K=100$ and $\Gamma=2$. 

  \subsection{{\tt PerBACCo}:  a general-relativistic 1D linear code}
  \label{sbsec:BERBACCO}
  
  The PerBACCo (PerturBAtive Constrained Code) general-relativistic
  linear code that we employ in this work is a development of the 
  one introduced in Refs.~\cite{Nagar:PhD,Nagar:2004av} and recently
  used in many studies~\cite{Nagar:2004ns,Bernuzzi:2008rq,Bernuzzi:2008fu}.
  This code is 1+1-dimensional and evolves, in the time domain, 
  nonspherical, matter and metric linear perturbations of a spherical star.
  The equations that are solved are obtained, after a multipolar decomposition
  of the linearized Einstein equations, as the static-background case in 
  the gauge-invariant and coordinate-independent formalism of 
  perturbations of spherically symmetric space-times developed in 
  Refs.~\cite{Gerlach:1979rw,Gerlach:1980tx,Seidel:1990xb,Gundlach:1999bt,MartinGarcia:2000ze}.
  We work explicitly in the Regge-Wheeler gauge. In this case, the full
  set of perturbation equations that we use is equivalent to that 
  of Refs~\cite{Allen:1997xj,Ruoff:2001ux}. 

  The focus of this work is on even-parity perturbations 
  only
  \footnote{The metric perturbations of a spherically 
    symmetric background space-time are divided in two classes, which are decoupled:
    the {\it even-parity} perturbation (also called {\it electric} because it is generated by the time 
    variation of the mass multipole moments of the source), which transform 
    as $(-1)^{\l}$ under a parity transformation, and the {\it odd-parity} perturbation
    (also called {\it magnetic} because it is generated by the current multipole
    moments), which transform as $(-1)^{\l+1}$.}
  .
  Let us recall that Ref.~\cite{Nagar:2004ns} showed how the even-parity 
  perturbation problem can be set up, and stably solved, using a constrained 
  formulation of the perturbation equations. These equations, as well as
  their numerical solution, have been discussed several times in the
  literature~\cite{Nagar:2004av,Nagar:2004ns,Bernuzzi:2008fu}. Notably,
  common practice is that
  (i) one elliptic equation, the 
  Hamiltonian constraint, namely Eq.~(7) of Ref.~\cite{Bernuzzi:2008fu}, is 
  solved to obtain the perturbed conformal factor, $k_{\lm}$; (ii) one
  hyperbolic equation, namely Eq.~(6) of Ref.~\cite{Bernuzzi:2008fu}, is used 
  (only inside the star) to 
  evolve the matter variable $H_{\lm}$ (\ie the perturbation of the relativistic 
  enthalpy); (iii) another hyperbolic equation, namely Eq.~(5) of
  Ref.~\cite{Bernuzzi:2008fu}, permits to obtain the nondiagonal, 
  gauge-invariant metric degree of freedom (the one actually associated with 
  gravitational radiation), $\chi_{\lm}$.
  After specification of initial data, the hyperbolic equations are solved 
  with standard, second-order-convergent-in-time-and-space, finite-differencing 
  algorithms (\eg leapfrog or Lax-Wendroff).
  Consistently, the elliptic equation is discretized at second order in
  space and reduced to a tridiagonal linear system, which is then solved by inversion.
  For any given multipole, $(\l,m)$, one solves the system of equations to
  obtain $\chi_{\lm}$ and $k_{\lm}$ as functions of time.
  Outside the star, one finally computes the Zerilli-Moncrief function as
  \begin{align}
  \label{eq:zerilli}
  \Psi^{(\rm e)}_{\lm} &=  \dfrac{2r(r-2M)}{\Lambda[(\Lambda -2)r +6M]}\nonumber\\
                       &\times \left[\chi_{\lm}- r \de_rk_{\lm} + \dfrac{r\Lambda+2M}{2(r-2M)} k_{\lm}\right],
  \end{align}
  where $M$ is the stellar mass and $\Lambda=\l(\l+1)$. This function is directly 
  connected to the $h_+$ and $h_\times$ gravitational-wave polarization amplitudes 
  [see Eq.~(\ref{eq:gi}) below] and it can be extracted from general-relativistic
  3D codes; for this reason it will be the main object of our interest in the
  forthcoming discussion. Note that Eq.~(\ref{eq:zerilli}) also defines our 
  normalization conventions and notation, that agree with those of Ref.~\cite{Nagar:2005ea}.

  \subsection{{\tt Cactus-Carpet-CCATIE-Whisky}:\\ 
              a general-relativistic 3D nonlinear code}
  \label{sbsec:fgrcode}

  We evolve a conformal-traceless ``$3+1$'' formulation of the Einstein
  equations~\cite{Nakamura87, Shibata95, Baumgarte99, Alcubierre99d}, in
  which the space-time is decomposed into three-dimensional spacelike
  slices, described by a metric $\gamma_{ij}$, its embedding in the full
  space-time, specified by the extrinsic curvature $K_{ij}$, and the gauge
  functions $\alpha$ (lapse) and $\beta^i$ (shift), that specify
  a coordinate frame (see Sec.~\ref{sbsec:gauges} for details on how we
  treat gauges and Ref.~\cite{York79} for a general description of the $3+1$
  split). The particular
  system which we evolve transforms the standard ADM variables as
  follows. The three-metric $\gamma_{ij}$ is conformally transformed via
  \begin{equation}
    \label{eq:def_g}
    \phi = \frac{1}{12}\ln \det \gamma_{ij}, \qquad
    \tilde{\gamma}_{ij} = e^{-4\phi} \gamma_{ij}
  \end{equation}
  and the conformal factor $\phi$ is evolved as an independent variable,
  whereas $\tilde{\gamma}_{ij}$ is subject to the constraint
  $\det \tilde{\gamma}_{ij} = 1$. The extrinsic curvature is
  subjected to the same conformal transformation and its trace
  $\tr K_{ij}$ is evolved as an independent variable. That is, in place of
  $K_{ij}$ we evolve
  \begin{equation}
    \label{eq:def_K}
    K \equiv \tr K_{ij} = g^{ij} K_{ij}, \qquad
    \tilde{A}_{ij} = e^{-4\phi} (K_{ij} - \frac{1}{3}\gamma_{ij} K),
  \end{equation}
  with $\tr\tilde{A}_{ij}=0$. Finally, new evolution variables
  \begin{equation}
    \label{eq:def_Gamma}
    \tilde{\Gamma}^i = \tilde{\gamma}^{jk}\tilde{\Gamma}^i_{jk}
  \end{equation}
  are introduced, defined in terms of the Christoffel symbols of
  the conformal three-metric.
  
  The Einstein equations specify a well-known set of evolution equations
  for the listed variables and are given by
  \begin{align}
    \label{eq:evolution}
    &\dt \tg_{ij} = -2 \alpha \tA_{ij}\,,  \\ \nonumber \\
    &\dt \phi = - \frac{1}{6} \alpha K\,, \\ \nonumber \\ 
    &\dt \tA_{ij} = e^{-4\phi} [ - D_i D_j \alpha 
      + \alpha (R_{ij} - 8 \pi S_{ij}) ]^{TF} \nonumber\\
    & \hskip 2.0cm + \alpha (K \tA_{ij} - 2 \tA_{ik} \tA^k{}_j), \\
    &\dt K  = - D^i D_i \alpha \nonumber \\
    & \hskip 2.0cm + \alpha \Big [\tA_{ij} \tA^{ij} + \frac{1}{3} K^2 + 
      4\pi (\rho_{_{\rm  ADM}}+S)\Big ], \\ \nonumber \\
    & 
    \partial_t \tG^i  = \tilde\gamma^{jk} \partial_j\partial_k \beta^i
    + \frac{1}{3} \tilde\gamma^{ij}  \partial_j\partial_k\beta^k
    + \beta^j\partial_j \tilde\Gamma^i
    - \Tilde\Gamma^j \partial_j \beta^i \nonumber \\
    & \hskip 1.0cm
    + \frac{2}{3} \tilde\Gamma^i \partial_j\beta^j 
    - 2 \tilde{A}^{ij} \partial_j\alpha
    + 2 \alpha ( 
    \tilde{\Gamma}^i{}_{jk} \tilde{A}^{jk} + 6 \tilde{A}^{ij}
    \partial_j \phi \nonumber\\
    & \hskip 1.0cm - \frac{2}{3} \tg^{ij} \partial_j K - 8 \pi \tg^{ij} S_j),
  \end{align}
  where $R_{ij}$ is the three-dimensional Ricci tensor. $D_i$ the
  covariant derivative associated with the three-metric $\gamma_{ij}$.
  ``TF'' indicates the trace-free part of tensor objects and $
  {\rho}_{_{\rm ADM}}$, $S_j$, and $S_{ij}$ are the matter source terms
  defined as
  \begin{align}
    \rho_{_{\rm ADM}}&\equiv n_\alpha n_\beta T^{\alpha\beta}, \nonumber \\ 
    S_i&\equiv -\gamma_{i\alpha}n_{\beta}T^{\alpha\beta}, \\
    S_{ij}&\equiv \gamma_{i\alpha}\gamma_{j\beta}T^{\alpha\beta}, \nonumber
  \end{align}
  where $n_\alpha\equiv (-\alpha,0,0,0)$ is the future-pointing four-vector
  orthonormal to the spacelike hypersurface and $T^{\alpha\beta}$ is the
  stress-energy tensor for a perfect fluid [{\it cf.} Eq. \ref{eq:stress-energy}]. 
  The Einstein equations also lead to a set of physical
  constraint equations that are satisfied within each spacelike slice:
  \begin{align}
    \label{eq:einstein_ham_constraint}
    \mathcal{H} &\equiv R^{(3)} + K^2 - K_{ij} K^{ij} - 16\pi\rho_{_{\rm ADM}} = 0, \\
    \label{eq:einstein_mom_constraints}
    \mathcal{M}^i &\equiv D_j(K^{ij} - \gamma^{ij}K) - 8\pi S^i = 0,
  \end{align}
  which are usually referred to as Hamiltonian and momentum constraints.
  Here $R^{(3)}=R_{ij} \gamma^{ij}$ is the Ricci scalar on a
  three-dimensional time-slice. Our specific choice of evolution
  variables introduces five additional constraints,
  \begin{align}
    \det \tilde{\gamma}_{ij} & = 1, 
    \label{eq:gamma_one}\\
  \tr \tilde{A}_{ij} & = 0,
  \label{eq:trace_free_A}\\
  \tilde{\Gamma}^i & = \tilde{\gamma}^{jk}\tilde{\Gamma}^i_{jk}.
  \label{eq:Gamma_def}
  \end{align}
  Our code actively enforces the algebraic
  constraints~(\ref{eq:gamma_one}) and~(\ref{eq:trace_free_A}). The
  remaining constraints, $\mathcal{H}$, $\mathcal{M}^i$,
  and~(\ref{eq:Gamma_def}), are not actively enforced and can be used
  as monitors of the accuracy of our numerical solution.
  See Ref.~\cite{Alcubierre02a} for a more comprehensive discussion of
  the above formalism.

  \subsubsection{Gauges}
  \label{sbsec:gauges}
  
  We specify the gauge in terms of the standard ADM lapse
  function $\alpha$, and shift vector $\beta^i$~\cite{misner73}.
  We evolve the lapse according to the ``$1+\log$'' slicing
  condition~\cite{Bona94b}:
  \begin{equation}
    \partial_t \alpha - \beta^i\partial_i\alpha 
    = -2 \alpha (K - K_0),
    \label{eq:one_plus_log}
  \end{equation}
  where $K_0$ is the initial value of the trace of the extrinsic
  curvature and
  equals zero for the maximally sliced initial data we consider here.
  The shift is evolved using the hyperbolic $\tilde{\Gamma}$-driver
  condition~\cite{Alcubierre02a},
  \begin{eqnarray}
    \label{shift_evol}
    \partial_t \beta^i - \beta^j \partial_j  \beta^i & = & \frac{3}{4} \alpha B^i\,,
    \\
    \partial_t B^i - \beta^j \partial_j B^i & = & \partial_t \tilde\Gamma^i 
    - \beta^j \partial_j \tilde\Gamma^i - \eta B^i\,,
  \end{eqnarray}
  where $\eta$ is a parameter which acts as a damping coefficient. The
  advection terms on the right-hand sides of these equations have been
  suggested in Refs.~\cite{Baker05a, Baker:2006mp, Koppitz-etal-2007aa}.
  
  All the equations discussed above are solved using the
  \texttt{CCATIE} code, a three-dimensional finite-differencing code
  based on the Cactus Computational Toolkit~\cite{Goodale02a}. A
  detailed presentation of the code and of its convergence properties
  have been recently presented in Ref.~\cite{Pollney:2007ss}.
  Mesh refinement is achieved through the Carpet code \cite{Schnetter-etal-03b}.

  \subsubsection{Evolution system for the matter}
  \label{sbsbsec:hd_eqs}
  
  We solve the general-relativistic hydrodynamics equations with the {\tt Whisky} 
  code~\cite{Baiotti:2008ra,Baiotti:2006wn,Baiotti:2006wm,Baiotti:2005vi,Baiotti04}.
  An important feature of the {\tt Whisky} code is the implementation of
  a \textit{flux-conservative} formulation of the hydrodynamics
  equations~\cite{Marti91,Banyuls97,Ibanez01}, in which the set of
  conservation equations for the stress-energy tensor $T^{\mu\nu}$ and
  for the matter current density $J^\mu=\rho u^\mu$, namely
  \begin{equation}
    \label{hydro eqs}
    \nabla_\mu T^{\mu\nu} = 0\;,\;\;\;\;\;\;
    \nabla_\mu J^\mu = 0\, ,
  \end{equation}
  is written in a hyperbolic, first-order and flux-conservative form of
  the type
  \begin{equation}
    \label{eq:consform1}
    \partial_t {\mathbf q} + 
    \partial_i {\mathbf f}^{(i)} ({\mathbf q}) = 
            {\mathbf s} ({\mathbf q})\ ,
  \end{equation}
  where ${\mathbf f}^{(i)} ({\mathbf q})$ and ${\mathbf s}({\mathbf q})$
  are the flux vectors and source terms, respectively~\cite{Font03}. Note
  that the right-hand side (the source terms) does not depend on derivatives 
  of the stress-energy tensor. Furthermore,
  while the system (\ref{eq:consform1}) is not strictly hyperbolic,
  strong hyperbolicity is recovered in a flat space-time, where ${\mathbf s}
  ({\mathbf q})=0$.
  
  As shown by Ref.~\cite{Banyuls97}, in order to write system
  (\ref{hydro eqs}) in the form of system (\ref{eq:consform1}), the
  \textit{primitive} hydrodynamical variables (\ie the
  rest-mass density $\rho$, the pressure $p$ measured in the
  rest-frame of the fluid, the fluid three-velocity $v^i$ measured by a local
  zero-angular momentum observer, the specific internal energy $\epsilon$
  and the Lorentz factor $W$) are mapped to the so-called \textit{conserved}
  variables \mbox{${\mathbf q} \equiv (D, S^i, \tau)$} via the relations
  \vspace{-0.2 cm}
  \begin{eqnarray}
    \label{eq:prim2con-d}
    D &\equiv& \sqrt{\gamma}W\rho\ , \\
    \label{eq:prim2con-s}
    S^i &\equiv& \sqrt{\gamma} \rho H W^2 v^i\ ,  \\
    \label{eq:prim2con-tau}
    \tau &\equiv& \sqrt{\gamma}\left( \rho H W^2 - p\right) - D\ .
  \end{eqnarray}
  Note that, in the case of a general EOS of the type $p=p(\rho,\epsilon)$ only five of the seven
  primitive variables are independent. Furthermore, if one adopts - as we do in the present work - a
  simpler isoentropic EOS of the type $p=p(\rho)$ where also the specific 
 energy ($\epsilon$) is fully determined by the rest-mass density ($\rho$), there is even one less independent variable. Namely
  Eq.~(\ref{eq:prim2con-tau}) becomes redundant and needs not be solved. No fundamental changes need
  being applied to the code, except that a simpler conversion scheme from conservative variables to
  primitive variables can be adopted~\cite{Baiotti2004thesis,Gourgoulhon2006}.

  In this approach, all variables ${\bf q}$ are represented on the
  numerical grid by cell-integral averages. The functions that the ${\bf q}$ represent are then {\it
  reconstructed} within each cell, usually by piecewise polynomials, in
  a way that preserves conservation of the variables ${\bf
  q}$~\cite{Toro99}. This operation produces two values at each cell boundary, which
  are then used as initial data for the local Riemann problems, whose (approximate)
  solution gives the fluxes through the cell boundaries. A method-of-lines
  approach~\cite{Toro99}, which reduces the partial differential
  equations~\eqref{eq:consform1} to a set of ordinary differential
  equations that can be evolved using standard numerical methods, such
  as Runge-Kutta or the iterative Cranck-Nicholson
  schemes~\cite{Teukolsky00,Leiler_Rezzolla06}, is used to update the
  equations in time (see Ref.~\cite{Baiotti03a} for further
  details). The {\tt Whisky} code implements several reconstruction
  methods, such as total-variation-diminishing (TVD) methods,
  essentially-non-oscillatory (ENO) methods~\cite{Harten87} and the
  piecewise parabolic method (PPM)~\cite{Colella84}. Also, a variety of
  approximate Riemann solvers can be used, starting from the
  Harten-Lax-van Leer-Einfeldt (HLLE) solver~\cite{Harten83}, over to
  the Roe solver~\cite{Roe81} and the Marquina flux
  formula~\cite{Aloy99b} (see Ref.~\cite{Baiotti03a,Baiotti04} for a more
  detailed discussion). 

  In this work we always use a global second-order accurate scheme, where time evolution is
  performed using the Iterative Cranck-Nicholson scheme with three substeps and with a
  Courant-Friedrichs-Lewy factor equal to 0.25. We always use the PPM method
  (that it is nominally 3rd-order accurate, but in actual simulations usually shows at best
  second-order accuracy) for the reconstruction and the Marquina formula for the approximate
  fluxes. The employed finite differencing for the space-time evolution with the {\tt
    CCATIE} code is fourth-order accurate. There are no particular reasons to prefer these schemes with
  respect to others used in the literature (like 3rd-order Runge-Kutta methods for time
  evolutions), however, since in this work we have focused on comparing gravitational-wave--extraction
  methods rather than time-evolution methods, we decided to use the old-fashioned iterative
  Cranck-Nicholson scheme.

\subsubsection{Treatment of the atmosphere}
\label{sec:atm-treatment}

  At least mathematically, the region outside our initial stellar
  models is assumed to be perfect vacuum. Independently of whether this
  represents a physically realistic description of a compact star, the
  vacuum represents a singular limit of the Eqs.~(\ref{eq:prim2con-d}-\ref{eq:prim2con-tau})
  and must be treated artificially. We have here followed a standard
  approach in computational fluid-dynamics and added a tenuous
  ``atmosphere'' filling the computational domain outside the star.

  We treat the atmosphere as a perfect fluid governed by the
  same polytropic EOS used for the bulk matter, but having a zero
  coordinate velocity. Furthermore, its rest-mass density is set to be several
  (6 in the present case) orders of magnitude smaller than the initial central
  rest-mass density. 
  
  The evolution of the hydrodynamical equations in grid-zones where
  the atmosphere is present is the same as the one used in the bulk of the
  flow. Furthermore, when the rest mass in a grid-zone falls below the
  threshold set by the atmosphere, that grid-zone is simply not updated in
  time and the values of its rest-mass density and velocity are set to those of 
  the atmosphere.

  \subsection{Gravitational-wave extraction in {\tt Cactus-Carpet-CCATIE-Whisky}}
  \label{sec:we}
  
  On a flat space-time, it is natural to express the waveform 
  as a multipolar expansion in spin-weighted spherical harmonics of 
  spin weight $s=-2$ as
  \begin{equation}
    \label{base}
    h_+ - {\rm i} h_\times = \sum_{\ell=2}^{\infty}\sum_{m=-\ell}^{\ell} h^{\ell m} {}_{-2}Y^{\ell m}(\theta,\phi) .
  \end{equation}
  The problem of gravitational-wave extraction out of a space-time computed numerically amounts 
  to computing, in a coordinate-independent way, the multipolar coefficients 
  $h^{\ell m}$.
  Two routes are commonly followed in numerical-relativity simulations of 
  astrophysical systems which do not involve matter (like binary black-hole
  coalescence).
  On one hand, one focuses on Weyl ``curvature'' waveforms~\cite{Campanelli:1998jv},
  by extracting from the numerical space-time the Newman-Penrose scalar
  $\psi_4$, which is related to the second time derivative of $(h_+,h_\times)$
  (see below).
  The metric waveform~\eqref{base} is then obtained from the curvature
  waveform via time integration. On the other hand, one can rely on 
  the Regge-Wheeler~\cite{RW57} and Zerilli~\cite{Zerilli:1970se}
  theory of metric perturbations of Schwarzschild space-time, 
  after recasting it in its gauge-invariant form according to Moncrief~\cite{Moncrief:1974am}.
  This allows to compute the metric waveform directly from the numerical space-time.
  See also Refs.~\cite{Nagar:2005ea,Sarbach:2001qq,Martel:2005ir} for reviews 
  and generalizations. Moreover, if matter is involved,
  it is also possible to calculate the 
  gravitational radiation emitted by the system
  by means of some (modified) Landau-Lifshitz quadrupole formula.
  The purpose of this section is to review the main elements of the
  three wave-extraction procedures, as an introduction to Sec.~\ref{sec:res},
  where waveforms obtained via the different methods will be compared
  and contrasted.

  \subsubsection{Wave-extraction via Newman-Penrose scalar $\psi_4$}
  \label{subs:NPform}

  The use of Weyl scalars for wave-extraction purposes has become very 
  common in numerical relativity and it has been successfully applied 
  in current binary--black-hole (see Ref.~\cite{Pretorius:2007nq} and 
  references therein), binary--neutron-star~\cite{Baiotti:2008ra} and 
  mixed-binary~\cite{Etienne:2007jg} simulations.

  Given a spatial hypersurface with timelike unit normal $n^\mu$ and given
  a spatial unit vector $r^\mu$ in the direction of the wave propagation,
  the standard definition of $\psi_4$ is the following component of the 
  Weyl curvature tensor $C_{\alpha\mu\beta\nu}$
  \begin{equation}
    \label{def_weyl}
    \psi_4 = - C_{\alpha\mu\beta\nu}\ell^\mu\ell^\nu\bar{m}^{\alpha}\bar{m}^\beta,
  \end{equation}
  where $\ell^\mu\equiv 1/\sqrt{2}(n^\mu-r^\mu)$  and $m^\mu$ is a complex
  null vector (such that $m^\mu\bar{m}_\mu=1$) that is orthogonal to $r^\mu$
  and $n^\mu$. This scalar can be identified with gravitational radiation
  if a suitable frame is chosen at the extraction radius.
  On a curved space-time there is considerable freedom in the choice of the
  vectors $r^\mu$ and $m^\mu$ and different researchers have made different
  choices, which are all equivalent in the $r\to\infty$ limit (see for 
  example~\cite{Nerozzi:2007ai} and references therein).
  We define an orthonormal basis in the 
  three-space $(\hat{e}_r,\hat{e}_{\theta},\hat{e_\phi})$, centered on the
  Cartesian origin and oriented with poles along the $z$-axis. The normal to
  the slice defines a timelike vector $\hat{e}_t$, from which we construct
  the null frame
  \begin{equation}
    l = \dfrac{1}{\sqrt{2}}(\hat{e}_t - \hat{e}_r),\;\;
    n = \dfrac{1}{\sqrt{2}}(\hat{e}_t + \hat{e}_r),\;\;
    m = \dfrac{1}{\sqrt{2}}(\hat{e}_{\theta}-\i e_{\phi}).
  \end{equation}
  We then calculate $\psi_4$ via a reformulation of Eq.~\eqref{def_weyl} in terms
  of ADM variables on the slice~\cite{Gunnarsen:1994sc},
  \begin{equation}
    \psi_4 = C_{ij}\bar{m}^i\bar{m}^j,
  \end{equation}
  where
  \begin{equation}
    C_{ij}\equiv R_{ij}-K K_{ij} + K_i^{\, k} K_{kj} - \i \epsilon_i^{\;kl} \nabla _l K_{jk}.
  \end{equation}
  The gravitational-wave polarization amplitudes $h_+$ and $h_\times$ are 
  related to $\psi_4$ by ~\cite{Teuk73}
  \begin{equation}
    \ddot{h}_+ - \i\ddot{h}_\times = \psi_4 .
  \end{equation}
  It is then convenient to expand $\psi_4$ in spin-weighted spherical harmonics
  of weight $s=-2$ as
  \begin{equation}
    \psi_4(t,r,\theta,\phi) = \sum_{\ell=2}^{\infty}\sum_{m=-\ell}^{\ell} 
    \psi_4^{\ell m}(t,r) {}_{-2}Y^{\ell m}(\theta,\phi)\, ,
  \end{equation}
  so that the relation between $\psi_4^{\ell m}$ and the 
  metric multipoles $h^{\ell m}$ becomes
  \begin{equation}
    \label{wave_psi4IN}
    \ddot{h}^{\ell m}(t,r) =  \psi_4^{\ell m}(t,r) \ .
  \end{equation}
 ${h}^{\ell m}(t,r)$ is then the double indefinite integral of $\psi_4^{\ell m}(t,r)$, which we
 numerically compute (after multiplying both sides by $r$) as
 \begin{equation}
   r\,  \tilde{h}^\lm(t,r) \equiv \int_{0}^{t}dt'\int_{0}^{t'}dt'' r\,
   \psi_4^\lm(t'',r) \ ,
  \end{equation}
  which results in
  \begin{eqnarray}
    r\,  h^{\ell m}(t,r) = r\tilde{h}^\lm(t,r)  + Q_0 + t \; Q_1   \ ,
   \end{eqnarray}
  where the integration constants $Q_0$ and $Q_1$ are explicitly written. They can be
  determined from the data themselves and their physical meanings are  $Q_0 = r  h^{\ell m}(0,r)$ and 
  $Q_1 = r\dot{h}^{\ell m}(0,r)$.

  This is not the end of the story yet. The equations discussed
  so far refer to a signal extracted at a {\it finite} value of
  $r$, while one is interested in computing $\psi_4^{\lm}$ at
  spatial infinity. It is imaginable that in the computed values of
  $\psi_4^{\lm}(t,r)$ there may be an offset, dependent on the extraction 
  radius; that is, $\psi_4^{\lm}$ at spatial infinity should be written as
  \begin{equation}
    r\psi_4^{\lm}(t) \equiv r\psi_4^{\lm}(t,r) + 2 Q_2(r)\,,
  \end{equation}
  where $\psi_4^{\lm}(t,r)$ is the scalar extracted at a finite radius $r$
  and $2 Q_2(r)$ is an \emph{offset} function, that takes into account
  (in an additive way) the effects of the extraction at a finite radius. The 
  time integration of this offset generates an additional term 
  that is quadratic in time, so that the final result for $r h^{\lm}(t)$ is
  \begin{equation}  
    r\, h^\lm(t) = r\,  \tilde{h}^\lm(t,r) + Q_0 + Q_1 t + Q_2(r) t^2 \; .
  \label{eq:psi4const}
  \end{equation}  
  The term $Q_2(r)$ should tend to zero when the extraction radius goes to 
  infinity. We checked that this is the case for the results of our simulations 
  (see Sec.~\ref{sbsc:waveforms} and Fig.~\ref{fig:Psie_Pert_u}).

  Various ways of fixing the two integration constants $Q_0$ and $Q_1$ have
  been discussed in the literature about coalescing binary
  black-hole systems~\cite{Berti:2007fi,Pollney:2007ss,Baker:2008mj,Damour:2008te}.
  In particular, in Appendix A of Ref.~\cite{Damour:2008te} the following
  procedure was proposed: (i) integrate the curvature waveform twice forward 
  in time (starting from $t=0$ and including the initial burst 
  of radiation due to the initial-data setup); (ii) Subtract the 
  linear-in-time offset present in there. This simple procedure led to an 
  accurate metric waveform which exhibited the correct circular polarization behavior. 
  A similar line was also followed in Ref.~\cite{Berti:2007fi}, where it was pointed
  out that in some situations (\eg close extraction radius, higher
  multipoles) one needs to subtract a general polynomial in $t$,
  consistently with our Eq.~\eqref{eq:psi4const}.

  \subsubsection{Abrahams-Price metric wave-extraction procedure}
  \label{ap_waves}
  
  The wave-extraction formalism based on the perturbation theory of 
  a Schwarzschild space-time was introduced
  by Abrahams and Price~\cite{Abrahams:1995gn} and subsequently employed 
  by many authors~\cite{Abrahams:1997ut,Camarda:1998wf,Allen:1998wy,Allen:1998rg}.

  The assumption underlying this extraction method is that, far from 
  the strong-field regions, the numerical space-time can be well approximated 
  as the sum of a spherically symmetric Schwarzschild ``background'' 
  $g_{\mu\nu}^{0}$ and a nonspherical perturbation $h_{\mu\nu}$. 
  Even if based on the gauge-invariant formulation of perturbations due 
  to Moncrief~\cite{Moncrief:1974am}, 
  the standard implementation~\cite{Abrahams:1995gn} of this approach 
  is done by fixing a coordinate system (Schwarzschild coordinates) for the background.
  As usual, the spherical symmetry
  \footnote{That is, the background 4-manifold $M$ can be written 
    as $M=M^2\times S^2$, where $M^2$ is a two-dimensional Lorentzian manifold
    and $S^2$ is the unit two-sphere.} 
  of $g_{\mu\nu}^0$ allows one to
  eliminate the dependence on the angles $(\theta,\phi)$ 
  by expanding 
  $h_{\mu\nu}$ in (tensor) spherical 
  harmonics, \ie seven even-parity and three odd-parity multipoles.
  The multipolar expansion explicitly reads
  \begin{equation}
    g_{\mu \nu}(t,r,\theta,\phi) = g_{\mu\nu}^{0} + \sum_{\ell=2}^{\infty}\sum_{m=-\ell}^{\ell} 
    \left[ \left(h_{\mu \nu}^{\ell m}\right)^{(\rm o)} + \left(h_{\mu \nu}^{\ell
        m}\right)^{(\rm e)} \right] \ .
  \end{equation}
  The metric multipoles $\left(h_{\mu\nu}^{\l m}\right)^{({\rm o/e})}$
  (and their derivatives) can be combined together in two gauge-invariant 
  master functions, the even-parity 
  (Zerilli-Moncrief) $\Psi^{(\rm e)}_{\ell m}$ [see Eq.~\eqref{eq:zerilli} above]
  and the odd-parity (Regge-Wheeler) $\Psi^{(\rm o)}_{\lm}$. These two master
  functions satisfy two decoupled wavelike equations with a
  potential\footnote{The equations are just approximately satisfied on the 
  extracted background.}.
  Finally, in a radiative coordinate system we have
  \begin{equation} 
    \label{eq:gi}
    h^{\ell m} = \dfrac{N_\l}{r}
    \left(\Psi^{(\rm e)}_{\ell m} + \i \Psi^{(\rm o)}_{\ell m}\right),
  \end{equation}
  where $N_\l=\sqrt{(\l+2)(\l+1)\l(\l-1)}$. 

  Note that the use of Schwarzschild coordinates for the background metric
  is not at all necessary and more general wave-extraction frameworks 
  exist. In particular, Sarbach and  Tiglio~\cite{Sarbach:2001qq}
  and Martel and Poisson~\cite{Martel:2005ir} have shown that there exists a 
  generalized formalism for perturbations that is not only
  gauge invariant (\ie invariant under infinitesimal coordinate
  transformation), but also {\it coordinate independent}, in the 
  sense that it is invariant under {\it finite} coordinate transformations
  of the $M^2$ Lorentzian submanifold of the background.
  Since in a numerical-relativity simulation the gauge depends on time, 
  one is {\it a priori} expecting that the gauge fixing of the background
  may introduce systematic errors. 
  For the odd-parity case, Ref.~\cite{Pazos:2006kz} has shown that this 
  is indeed the case for the particular physical setting represented by
  the scattering of a Gaussian pulse of gravitational waves on a 
  Schwarzschild black hole in Kerr-Schild coordinates (see
  Ref.~\cite{Korobkin:2008ji} for the even-parity case).
  In this work we present results obtained using the
  ``standard'' Moncrief formalism. A comprehensive discussion of
  results obtained via the generalized formalism will be presented 
  elsewhere~\cite{bernuzzi_prep}.

  \subsubsection{Landau-Lifshitz quadrupole-type formula}

  In the presence of matter, it is sometimes convenient to extract gravitational
  waves using also some kind of (improved) Landau-Lifshitz ``quadrupole''
  formula. Although this formula is not gauge invariant, this route has 
  been followed by many authors with different degrees of 
  sophistication~\cite{Finn:1990,Dimmelmeier:2002bm,
    Shibata:2003aw,Shibata:2004kb,Shibata:2005ss}, 
  to give well approximated waveforms~\cite{Shibata:2003aw}.  
  For the sake of completeness, let us review how this quadrupole
  formula came into being, as the first contribution in a multipolar expansion,
  and let us express it in the convenient form of $h^{\ell m}$, as
  outlined above. The basic reference of the formalism is 
  a review by Thorne~\cite{Thorne:1980ru}; most of the useful formulas of this review
  have been collected by Kidder~\cite{Kidder:2007rt}, who condenses and
  summarizes the gravitational-wave--generation formalism developed in
  Refs.~\cite{Blanchet:1986sp, Blanchet:1987wq}.
  
  Following Ref.~\cite{Kidder:2007rt}, we recall that Eq.~(\ref{eq:gi})
  can be derived in all generality by (i) decomposing the asymptotic 
  waveform $h_{ij}^{\rm TT}$ into two sets of symmetric trace-free (STF) 
  radiative multipole moments (to be related later to the matter multipole 
  moment of the source in the near-zone) called ${\cal U}_L$ and 
  ${\cal V}_L$, where a capital letter for an 
  index denotes a multi-index (\ie, ${\cal U}_L={\U}_{i_1 i_2\dots i_\l}$); 
  (ii) projecting the STF-decomposed $h_{ij}^{\rm TT}$ 
  along an orthonormal triad that corresponds to that of the 
  spherical coordinate system.
  In the same notation of Ref.~\cite{Kidder:2007rt}, 
  Eq.~(\ref{eq:gi}) reads
  \begin{equation}
    h^{\ell m} = \dfrac{1}{\sqrt{2} r}\left(U^{\ell m} - \i V^{\ell m}\right)\,,
  \end{equation}
  where the mass multipole moments $U^{\ell m}$ and current multipole 
  moments $V^{\ell m}$ are related to their STF counterparts by
  \begin{align}
    U^{\ell m} &= \dfrac{16\pi}{(2\ell +1)!!}
    \sqrt{\dfrac{(\ell+1)(\ell+2)}{2\ell (\ell-1)}}{\cal U}_L {\cal Y}^{\ell m*}_L ,\\
    V^{\ell m} &= -\dfrac{32\pi\ell}{(2\ell +1)!!}
    \sqrt{\dfrac{\ell +2}{2\ell(\ell+1)(\ell-1)}}{\cal V}_L {\cal Y}^{\ell m*}_L ,
  \end{align}
  where ${\cal Y}^{\ell m}_{L}$ are the STF spherical harmonics.
  These functions form a basis of the of the $(2\ell +1)$-dimensional 
  vector space of STF ${\ell}$-tensors; they are related to the scalar 
  spherical harmonics by
  \begin{equation}
    Y^{\ell m} = \Y^{\l m}_L N_L,
  \end{equation}
  where $N_i$ is a component of the unit radial vector. 
  The expanded form of the STF $\Y^{\l m}_L$ is given in 
  Refs.~\cite{Pirani:1964,Thorne:1980ru} (see also Eq.~(A6a) 
  of Ref.~\cite{Blanchet:1986sp}).
  In the post-Newtonian (PN) wave-generation formalism of 
  Refs.~\cite{Blanchet:1986sp,Blanchet:1987wq}, one can 
  relate in a systematic manner the radiative multipole 
  moments (${\cal U}_L,{\cal V}_L$) to a set of six STF 
  source moments ($\I_L,\J_L,\W_L,\X_L,\Y_L,\Z_L$), which 
  can be computed  from the stress-energy pseudotensor of 
  the matter and of the gravitational field of the source. 
  A set of two canonical source moments $({\M}_L, {\S}_L)$ 
  can be computed as an intermediate step between the source moments 
  and the radiative moment. Two of the source moments, the mass
  moments $\I_L$ and the current moment $\J_L$ are dominant, while 
  the others only make a contribution starting at 2.5~PN order
  and we neglect them here. In a first approximation (\ie neglecting 
  the nonlinear ``tail-interactions'' as well as higher-order nonlinear 
  interactions), the $L$-th radiative moment is given by the $\ell$-th 
  time derivative of the canonical moments as
  \begin{align}
    \label{eq:multipoles}
    \U_L &\equiv {\I}^{(\ell)}_L + O(\varepsilon^{5/2}) ,\\
    \V_L &\equiv {\J}^{(\ell)}_L + O(\varepsilon^{5/2}),
  \end{align}
  where $\varepsilon\sim (v/c)^2$ indicates some PN ordering parameter of
  the system. As a result, the computation of $U_{\l m}$ and $V_{\l m}$ 
  is straightforward. As an example (that will be used in the following), 
  let us focus on the $\l=2$ moments of a general astrophysical system
  with equatorial symmetry. In this case, the $(2,1)$ moment is  
  purely odd-parity, while the $(2,0)$ and $(2,2)$ are purely even-parity.
  Straightforward application of what we have reviewed so far gives
  \begin{align}
    \label{eq:multipoles_20}
    h^{20} &=\dfrac{1}{r}\sqrt{\dfrac{24\pi}{5}}\left(\ddot{\I}_{zz}
    - \dfrac{1}{3}{\rm Tr}{(\ddot{\I})}\right), \\
    \label{eq:multipoles_21}
    h^{21}&=-\dfrac{\i}{r}\sqrt{\dfrac{128\pi}{45}}\left(\ddot{\J}_{xz} - \i \ddot{\J}_{yz}\right),\\
    \label{eq:multipoles_22}
    h^{22} &=\dfrac{1}{r}\sqrt{\dfrac{4\pi}{5}}\left(\ddot{\I}_{xx}-2\i\ddot{\I}_{xy}-\ddot{\I}_{yy}\right).
  \end{align}
  In the harmonic gauge, in the case of small velocity and negligible internal
  stresses (\ie in the Newtonian limit) one has $\I_{ij}=\int d^3 x \rho x_i x_j$
  and $\J_{ij} = \int d^3x\rho\varepsilon_{abi}x_j x_a v^b$.
  The 1~PN corrections to the mass quadrupole have been computed in
  Ref.~\cite{Blanchet:1989fg}. Recently, Ref.~\cite{CerdaDuran:2004vn}
  included 1~PN correction, using an effective 1~PN quadrupole momentum,
  in the gravitational-wave--extraction procedure from supernova core-collapse simulations.
  As a complementary approach, Ref.~\cite{Shibata:2003aw} proposed to
  ``effectively'' take into account possible general-relativistic corrections 
  by inserting in Eqs.~(\ref{eq:multipoles_20}-\ref{eq:multipoles_22}) the 
  following effective ``quadrupole moment'' defined in terms of 
  the ``coordinate rest-mass density'' $\rho_*\equiv \alpha\sqrt{\gamma}u^0\rho$,
  \begin{equation}
    \label{eq:SQF3}
    \I_{ij} = \int d^3 x\rho_* x_i x_j\,.
  \end{equation}
  This presents some very useful properties: (i) it is of 
  simple implementation and (ii) from the continuity 
  equation $\de_t\rho_*+\de_i(\rho_*v^i)=0$, one can analytically 
  compute the first time-derivative of the quadrupole moment, 
  so that only one numerical time-derivative needs to be evaluated.
  The last property is extremely important, in fact, on data computed 
  via a second-order accurate numerical scheme it is not possible to
  calculate noise-free third derivatives, which are needed for the
  gravitational-wave luminosity.
  The accuracy of a scheme based on Eq.~(\ref{eq:SQF3}) has been tested 
  in Ref.~\cite{Shibata:2003aw} in the case of neutron-star
  oscillations and was subsequently used by various authors 
  to estimate the gravitational-wave emission in other physical 
  scenarios. See for example Refs.~\cite{Shibata:2004nv,Shibata:2004kb,Baiotti:2006wn,Dimmelmeier:2007ui}.
  In order to get some more insight on the accuracy of possible
  ``generalized'' standard quadrupole formulas (SQFs formulas), we have tried the strategy
  exploited in Ref.~\cite{Nagar:2005cj}, namely to test some pragmatic
  modifications of the quadrupole formula and to check which one
  is closer to the actual gravitational waveform. 
  In practice, we start with a sort of generalized ``quadrupole moment''
  of the form
  \begin{equation}
    \label{eq:quadrupole}
    \I_{ij}[\DENSq] \equiv \int d^3 x \DENSq x_i x_j\, ,
  \end{equation}
  where now, instead of the ``Matter density'', we use
  the following generalized effective densities $\DENSq$:
  \begin{align}
    \label{sqf}
    \mbox{SQF } &\;\;\; \DENSq:=\rho\,,\\
    \mbox{SQF1} &\;\;\; \DENSq:=\alpha^2\sqrt{\gamma}T^{00}\,,\\
    \mbox{SQF2} &\;\;\; \DENSq:=\sqrt{\gamma}W\rho\,,\\
    \label{sqf3}
    \mbox{SQF3} &\;\;\; \DENSq:=u^0\rho=\frac{W}{\alpha}\rho\,.
  \end{align}
  We do not think that any of the ``quadrupole formulas'' obtained using these
  generalized quadrupole moments should be considered better than the others. 
  Note that none of them is gauge invariant and, indeed, the outcome will
  change if one is considering isotropic or Schwarzschild-like coordinates.
  These formulas were widely used in the literature and the main purpose of 
  the comparison among Eqs.~(\ref{sqf}-\ref{sqf3}) is to give an idea
  of the kind of information that can be safely assessed using them.
  We will comment more on that in the discussion in the 
  following Sec.~\ref{sbsc:WaveQuad}.
  
  \section{Initial data}
  \label{sbsc:init}
  As a representative model for a neutron star,
  we choose a model described by a polytropic EOS [Eq.~\ref{eq:EOS}]
  with $\Gamma=2$, $K=100$, central rest-mass density $\rho_c=1.28\times10^{-3}$ 
  and so with rest mass $M\simeq 1.4$. 
  This model has been widely used in the literature and it is known
  as model A0 in Ref.~\cite{Stergioulas:2003ep}. 
  Some of its equilibrium properties are listed in Table~\ref{tab:starmodel}.
  
  \subsection{Fluid-perturbation setup}
  \label{sbsbsc:fluid}
  
  In both the linear and nonlinear codes, setting up the initial data amounts
  to (i) solving the Tolman-Oppenheimer-Volkov (TOV) equations to construct the 
  equilibrium configuration; (ii) fixing an axisymmetric pressure perturbation; 
  and (iii) solving the linearized constraints for the metric perturbations.
  We rewrite the perturbative equations in terms of enthalpy perturbations 
  because it is more convenient.
    
  We set up the initial pressure perturbation as an axisymmetric 
  multipole:
  \begin{equation}
    \label{eq:pressure_perturbation}
    \delta p (r,\theta) \equiv (p+e)H_{\ell 0}(r)Y_{\ell 0}(\theta) \ ,
  \end{equation}
  and then one is free to specify a profile for the relativistic
  enthalpy $H_{\ell 0}(r)$. Actually we limit our study to $\l=2$ (quadrupole) perturbations.
  Since we aim at a comparison between waveforms and not at exploring the
  physics of the process of neutron-star oscillations, for our purpose the best 
  system is represented by a star oscillating precisely at one frequency, 
  \ie such that $H_{\ell 0}(r)$ corresponds to an eigenfunction of the star. 
  \begin{table}[t]
    \caption{\label{tab:starmodel} Equilibrium properties of model A0. From
      left to right the columns report: central rest-mass density, central total 
      energy density, gravitational mass, radius, compactness.}
    \begin{ruledtabular}
      \begin{tabular}{cccccc}
	Name  &   $\rho_c$ & $e_c$ & $M$ & $R$ & $M/R$\cr
	\hline
	A0 & $1.28\times10^{-3}$ & $1.44\times10^{-3}$ & $1.40$ & $9.57$ & $0.15$\cr
    \end{tabular}
    \end{ruledtabular}
  \end{table}
  We set a profile of $H_{\ell 0}(r)$ that excites, mostly, the 
  $f$ mode of the star (with a small contribution from the first overtone). 
  In general, as suggested in Ref.~\cite{Nagar:2004av}, 
  an ``approximate eigenfunction'' for a given fluid mode can be 
  given by setting
  \begin{equation}
    \label{eq:forH}
    H_{\ell 0}=\lambda\sin\left[\dfrac{(n+1)\pi r}{2R}\right]\,,
  \end{equation}
  where $n$ is an integer controlling the number of nodes of $H_{\ell 0}(r)$, 
  $\lambda$ is the amplitude of the perturbation and $R$ is
  the radius of the star in Schwarzschild coordinates. The case $n=0$ has no
  nodes (\ie no zeros) for $0<r\leq R$; as a result, the $f$ mode is 
  predominantly triggered (as in Ref.~\cite{Nagar:2004av}) and the $p$-mode 
  contribution is negligible. If $n=1$, the $f$ mode is still dominant, 
  but a nonnegligible contribution of the $p_1$ mode is present. If
  $n=2$, in addition to the fundamental and the first pressure modes also 
  the $p_2$ mode is clearly present in the signal.
  For higher values of $n$ more and more overtones are excited.
  \begin{table}[t]
    \caption{\label{tab:ID:depres} From the enthalpy perturbation to the relative magnitude
      of the pressure perturbation for $n=0$, $\ell=2$, and $m=0$ (see
      Fig.~\ref{fig:ID:depres}). The minimum pressure perturbation
      occurs at some value of $r$ on the $xy$ plane ($\theta=\pi/2$), while the 
      maximum pressure perturbation is found at some value of $r$ on the $z$ axis ($\theta=0$).}
    \begin{ruledtabular}
      \begin{tabular}{llcc}
        Name  & $\lambda$  & min($\delta p/p_c$) & max($\delta p/p_c$)\\
        \hline
        \hline
        $\lambda0$ & 0.001 & -0.00125 & 0.00251\\
        $\lambda1$ & 0.01  & -0.01253 & 0.02506\\
        $\lambda2$ & 0.05  & -0.06266 & 0.12533\\
        $\lambda3$ & 0.1   & -0.12533 & 0.25067\\
      \end{tabular}
    \end{ruledtabular}
  \end{table}
  \begin{figure}[t]
    \begin{center}
      \includegraphics[width=85 mm]{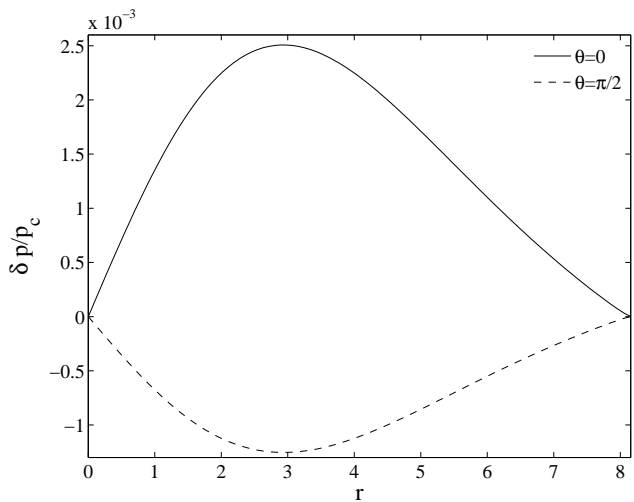}\\
      \caption{ \label{fig:ID:depres} 
        Profile of the relative pressure perturbation
        $\delta p/p_c$ [computed on the $z$-axis ($\theta=0$) and on the $xy$-plane
        ($\theta=\pi/2$)] obtained from Eqs.~(\ref{eq:pressure_perturbation}-\ref{eq:forH}) 
        with $n=0$ and $\lambda=\lambda0$.}
    \end{center}
  \end{figure}

  In the following, we use the same setup, Eq.~(\ref{eq:forH}), to provide initial data
  in both the linear and nonlinear codes.
  Correspondingly, the computation of $\delta p$ is needed to get
  a handle on the magnitude of the deviation from sphericity. The best 
  indicator is given by the ratio $\delta p/p_c$, where $p_c$ is
  the central pressure of the star. Fig.~\ref{fig:ID:depres} displays 
  the profile of $\delta p/p_c$, at the pole and at the equator, as 
  a function of the Schwarzschild radial coordinate $r$ for $\lambda=\lambda0=0.001$.
  For simplicity, we consider only $n=0$ perturbations, with four values 
  for the amplitude, namely $\lambda=[0.001,~0.01,~0.05,~0.1]$, in order to see,
  in the 3D code, how the transition from linear to nonlinear regime
  occurs. Maxima and minima of the initial pressure perturbation for the different 
  values of the initial perturbation amplitude $\lambda$ can be found in Table~\ref{tab:ID:depres}.

  \subsection{Metric-perturbation setup: the 1D linear code}
  \label{sbsbsc:1dmetric}
  
  Let us turn now to discuss the implementation of Eq.~(\ref{eq:forH}) 
  in the two codes and the corresponding treatment of the related 
  initial metric perturbation. 
  As discussed in Refs.~\cite{Gundlach:1999bt,Seidel:1990xb}, 
  the even-parity metric  perturbation of a general (nonstatic) spherically 
  symmetric space-time is described by 3 degrees of 
  freedom ($k_{\l m},\chi_{\l m}$ and $\psi_{\l m}$) that are the solution of 
  three coupled partial differential equations. 
  On the static TOV background, only $k_{\l m}$ and $\chi_{\l m}$
  are independent degrees of freedom of the gravitational field,
  and their evolution equations are decoupled from that of $\psi_{\l m}$,
  which can be obtained at every time-step once $\chi_{\l m}$ and $k_{\lm}$
  are known. The recovery of $\psi_{\lm}$ from $\chi_{\lm}$ and $k_{\lm}$,
  that is needed in the 3D case, will be explicitly discussed in 
  Sec.~\ref{sbsbsc:3dmetric} below. By contrast, for the 1D implementation
  one only needs to specify initial data for 
  $\chi_{\l m}$, $k_{\l m}$, and their time derivatives. This is 
  accomplished by solving the constraints under a number
  of assumptions related to the physics that we want to investigate.
  First of all, we consider only axisymmetric perturbations ($m=0$) 
  and we restrict ourselves to the dominant quadrupole mode ($\l=2$). 
  Then, since in this work 
  we are not interested in $w$-mode excitation, we impose the 
  conformally flat condition ($\chi_{20}=0$) (see Refs.~\cite{Nagar:2004ns,Bernuzzi:2008fu} 
  for details). With These hypotheses, we solve the Hamiltonian
  constraint, namely Eq.(7) of Ref.~\cite{Bernuzzi:2008fu}, for $k_{20}$.
  This is done on a grid $r\in [0,r_{\rm max}]$, with $r_{\rm max}\gg R$
  and with boundary condition $k_{20}=0$ at $r=0$ and at $r=r_{\rm max}$.
  We impose $\dot{k}_{20}=\dot{\chi}_{20}=0$ for simplicity, but 
  we are aware that this is inconsistent with the condition
  that $H_{20}\neq 0$ initially and thus the momentum constraints
  should also be solved. However, since the effect is a small
  initial transient in the waveforms that quickly washes out 
  before the quasiharmonic oscillation triggered by the perturbation $H_{20}$ 
  sets in, we have decided to maintain the initial-data setup simple.
  Figure~\ref{fig:ID:k_psie} synthesizes the information about the initial
  data. The top panel shows (as a solid line) the profile of $k_{20}$ 
  (versus Schwarzschild radius)
  corresponding to the perturbation $\lambda=\lambda0$ of Table~\ref{tab:ID:depres}; 
  the bottom panel shows 
  (as a solid line) the initial profile of the Zerilli-Moncrief 
  function $\Psi^{(\rm e)}_{20}$ outside the star.
  
  \subsection{Metric-perturbation setup: the 3D nonlinear code}
  \label{sbsbsc:3dmetric}
    
  In the 3D code we setup the same kind of initial condition 
  as in the 1D code, but the procedure is more complicated as
  one needs to reconstruct the full 3D metric on the Cartesian
  grid. In addition, the main difference with respect to the
  1D case is that the perturbative constraints are expressed
  using a radial {\it isotropic} coordinate $\bar{r}$ instead 
  of the Schwarzschild-like radial coordinate $r$. This is done 
  because $\bar{r}$ is naturally connected to the Cartesian 
  coordinates in which the code is expressed, \ie $\bar{r}=\sqrt{x^2+y^2+z^2}$. 
  The initialization of the metric in the 3D case has to follow
  four main steps: (i) The perturbative constraints are solved, 
  (ii) The multipolar metric components are added to the
  unperturbed background TOV metric; (iii) The resulting metric is written
  in Cartesian coordinates; (iv) It is interpolated 
  on the Cartesian grid. 

  Let us then recall some useful formulas. At the background level, 
  the TOV metric in isotropic coordinates reads
  \begin{equation}
    \label{eq:backg}
    ds^2_0 = -e^{2a} dt^2 + e^{2b}\left(d\r^2 + \r^2d\Omega^2\right) \ ,
  \end{equation}
  where $d\Omega=d\theta^2+\sin^2d\phi^2$.
  The relations between the Schwarzschild and isotropic radial coordinates
  in the exterior are given by
  \begin{align}
    \label{r_from_barr}
    r &= \r\left( 1+\frac{M}{2\r} \right)^2 \ ,\\ 
    \r &= \half\left( \sqrt{r^2 -2Mr} +r -M \right) \ ,
  \end{align}
  and in the interior by
  \begin{align}
    \label{r_from_barr_out}
    r &= \r e^{2 b(r)} \ ,\\
    \r &= C r\exp{\left[ \int^r_0 dx
    \frac{1-\sqrt{1-2m(x)/x}}{x\sqrt{1-2m(x)/x}}  \right]} \ ,
  \end{align}
  where
  \begin{align}
    C &=\frac{1}{2R}\left( \sqrt{R^2-2MR} +R -M \right) \nonumber \\
    &\times\exp{\left[-\int_0^Rdx\frac{1-\sqrt{1-2m(x)/x}}{x\sqrt{1-2m(x)/x}} \right]}.
  \end{align}
  In terms of the isotropic radius, the perturbative Hamiltonian constraint
  explicitly reads
  \begin{align}
    \label{iso_constraint}
    &k^{\lm}_{,\r \r} +
    \dfrac{k^{\lm}_{,\r}}{\r}\left[1+\left(1-\dfrac{2m(r)}{r}\right)^{1/2}\right]\nonumber\\
    &
    + e^{2b}\left(8\pi e -\dfrac{\Lambda}{r^2}\right) k^{\lm}
    -\dfrac{1}{\r}\left(1-\dfrac{2 m(r)}{r}\right)^{1/2}\!\!\!\chi_{,\r}^{\lm} \nonumber\\
   &- e^{2b}\left(\dfrac{\Lambda+2}{2r^2}-8\pi e\right)\chi^{\lm}
   =-\dfrac{8\pi(p+e)H^{\lm}}{C_s^2} \,,
  \end{align}
  where $r\equiv r(\r)$ according to Eq.~\eqref{r_from_barr} and~\eqref{r_from_barr_out}.

  \begin{figure}[t]
    \begin{flushright}
      \includegraphics[width=83 mm]{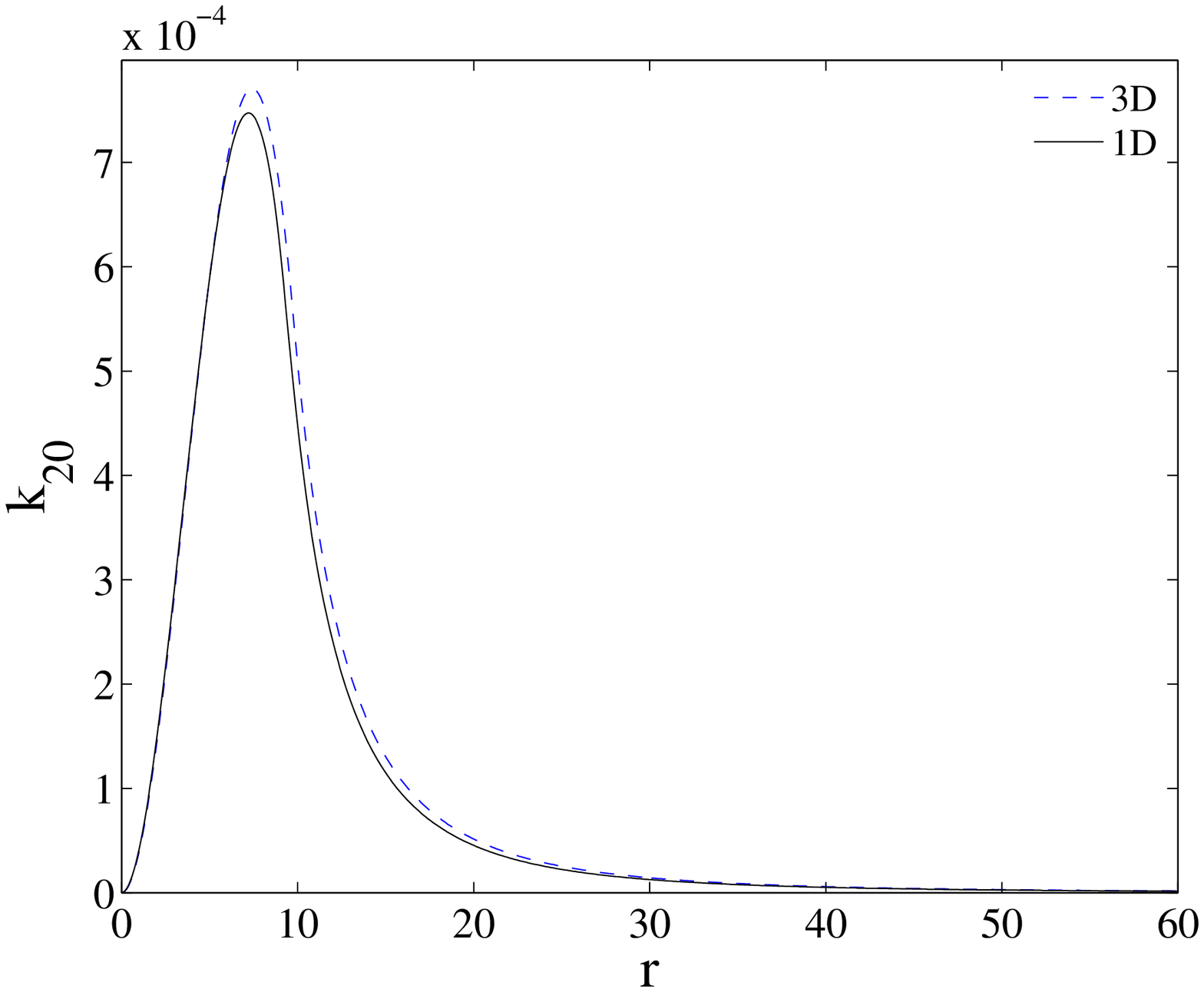}\\
      \includegraphics[width=86 mm]{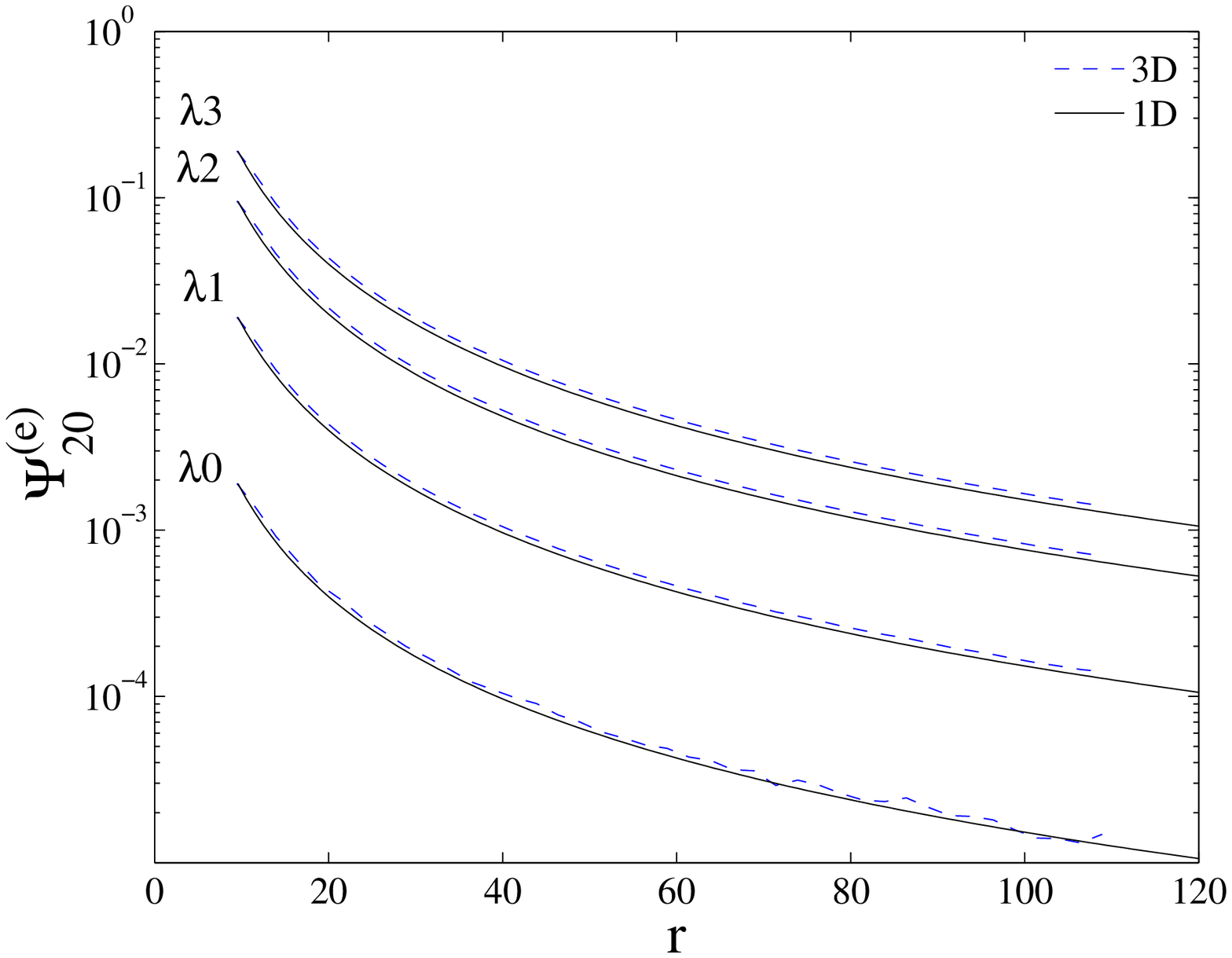}
    \end{flushright}
    \vspace{-4mm}
      \caption{\label{fig:ID:k_psie} (color online)
        Initial-data setup in the 1D and 3D
	codes. {\bf Top panel}: Profiles of the $k_{20}$
	multipole at $t=0$ versus the Schwarzschild radial coordinate r, 
	obtained from the solution of the perturbative Hamiltonian constraint 
	with $\lambda=\lambda0$ and $n=0$. {\bf Bottom panel}: Profiles of $\Psi^{(\rm e)}_{20}$ at 
	$t=0$ versus the Schwarzschild radial coordinate $r$, for different 
	values of the initial perturbation $\lambda$.
        Both panels compare the results of the computations of the 1D and 3D codes.}
  \end{figure}
  After solving the TOV equations, we choose a profile
  for $H_{\lm}$, impose the conformal-flatness condition $\chi_\lm=0$, and solve
  Eq.~\eqref{iso_constraint} for $k_{\lm}$. As we discussed in the previous 
  section, one also needs to impose on $\psi_\lm$ some condition that can 
  be regarded as an initial gauge condition. Then $(\chi_{\lm},k_{\lm},\psi_{\lm})$
  must be inserted in the explicit expression of the (even-parity) metric perturbation
  \begin{align}
    \label{eq:delta_gmunu}
    \delta s^2_\lm &= \big\{ (\chi_\lm+k_\lm)e^{2a}dt^2 - 2\psi_\lm
    e^{a+b}dtd\bar{r}  \nonumber\\
    & +e^{2b}\left[(\chi_\lm+k_\lm)d\r^2+\r^2k_\lm d\Omega\right]\big\}Y_{\l m}  \ .
  \end{align}
  In the absence of azimuthal and tangential velocity perturbations,
  in Schwarzschild coordinates and in the Regge-Wheeler gauge, 
  from the momentum-constraint equation, namely Eq.~(95) of 
  Ref.~\cite{Gundlach:1999bt}, one obtains
  \begin{equation}
    \psi_{\lm,r} = -\dfrac{2\left[m(r)+4\pi r^3 p(r)\right]}{r-2m(r)}\psi \,,
  \end{equation}
  which, once solved, gives
  \begin{equation}
    \psi_\lm(r) = \tilde{C} \exp\left[-\int_0^r dx \dfrac{2\left(m(x)+4\pi
    x^3 p(x)\right)}{x-2 m(x)}\right] .
  \end{equation}
  The requirement $\psi_\lm\to 0$ for large $r$, like for $k_\lm$,
  implies $\tilde{C}=0$; therefore, the metric perturbation is given by
  Eq.~\eqref{eq:delta_gmunu} with $\psi_\lm=\chi_\lm=0$.
  The full metric in isotropic coordinates 
  is obtained as $ds^2 = ds_0^2 + \delta s_{\lm}^2$. 
  This metric is transformed to Cartesian coordinates and then
  it is linearly interpolated onto the 
  Cartesian grid used to solve the coupled Einstein-matter equations
  numerically. To ensure a correct implementation of the boundary
  conditions (\ie $k_{\lm}\to 0$ when $\bar{r}\to \infty$), the
  isotropic radial grid used to solve Eq.~\eqref{iso_constraint} 
  is much larger ($\bar{r}\sim 3000$) than the corresponding 
  Cartesian grid ($\bar{r}\sim 208$) and the spacing is 
  much smaller.

  One proceeds similarly for the matter perturbation: 
  From a given profile for $H_\lm(r)$, the pressure perturbation  
  $\delta p(r,\theta)$ is computed and from this the total pressure
  is given by $p+\delta p$. This is interpolated on the Cartesian grid
  to finally obtain the vector of the conserved hydrodynamics variables
  $(D,S^i,\tau)$.
  
  The consistency of the initial-data setup procedure in both the 
  {\tt PerBACCo} 1D linear code and in the 
  {\tt Cactus-Carpet-CCATIE-Whisky} 3D nonlinear code is highlighted 
  in Fig.~\ref{fig:ID:k_psie}. The top panel of the figure compares 
  the profiles of $k_{20}$ in the 1D case  (solid line) and in the 3D case 
  (dashed line) for $\lambda=\lambda0$ and $n=0$. 
  The small differences are related to a slightly different location 
  of the star surface in the two setups and to the different resolution of the grids. 
  The bottom panel of 
  Figure~\ref{fig:ID:k_psie} contrasts the Zerilli-Moncrief functions 
  $\Psi^{(\rm e)}_{20}$ from the 1D code (solid lines) with those extracted
  (at $t=0$) from the numerical 3D metric (dashed lines). For all initial conditions, the curves
  show good consistency. 
  
  \section{Results}
  \label{sec:res}
  
  The presentation of our results is organized in the following way.
  In Sec.~\ref{sbsec:radial} we focus first on radial
  oscillations, that are always present due to numerical discretization
  error. Then we concentrate on nonradial 
  stellar oscillations and we compare the 1D and 
  3D metric waveforms (Sec.~\ref{sbsc:waveforms}) and curvature waveforms (Sec.~\ref{sbsc:waveformsPsi4}).
  In Sec.~\ref{junk_RWZ} we discuss
  advantages and disadvantages of these two wave-extraction techniques.
  Finally, we discuss the use of quadrupole-type 
  formulas in Sec.~\ref{sbsc:WaveQuad}, while Sec.~\ref{sbsc:nnlin}
  is devoted to the analysis of nonlinear couplings between oscillation modes.
  
  \subsection{Radial oscillations}
  \label{sbsec:radial}
  
  The unperturbed configuration A0 has been stably evolved 
  for about 20 ms. The numerical 3D grid used for this simulation
  is composed of two concentric cubic boxes with limits $[-32,\,32]$ and $[-16,\,16]$
  in all the three Cartesian directions. The boxes have resolutions
  $\Delta_{xyz}=0.5$ and $0.25$ respectively;
  bitant symmetry, \ie the $z<0$ domain is copied from the $z>0$ domain instead of being evolved, 
  was imposed as a boundary condition in order to save computational time.
  
  The truncation errors of the numerical scheme 
  trigger (physical) radial oscillations of (mainly) the fundamental 
  mode $F$ and the first overtones. We have checked that these 
  frequencies agree with those computed evolving the radial 
  pulsation equation with the perturbative code. This comparison
  is shown in Table~\ref{tab:radialmode}. We note in passing that 
  our numbers are in perfect agreement with those of Table~I 
  of Ref.~\cite{Font:2001ew}.
  
  \begin{table}[t]
    \caption{\label{tab:radialmode}
      Frequencies of the fundamental radial mode of model A0.}
    \begin{ruledtabular}
      \begin{tabular}{lccc}
	n & Pert.[Hz] & 3D [Hz] & Diff. [\%]\\
	\hline
	0 & 1462 & 1466 & 0.3\\
	1 & 3938 & 3935 & 0.1\\
	2 & 5928 & 5978 & 0.8\\
      \end{tabular}
    \end{ruledtabular}
  \end{table}
  
  As a further check, the entire sequence of uniformly rotating models
  with mass $M=1.4$ and nonrotating limit A0 has been evolved.
  Simulations were done with a cubic grid with limits $[-32,\,32]$ in each direction,
  and uniformly spaced with grid spacing $\Delta_{xyz}=0.5$. 
  As before, we have imposed bitant symmetry. The sequence of initial models has been computed 
  by means of the version of the {\tt RNS} code \cite{Stergioulas:1994ea} implemented
  in {\tt Whisky}. For the equilibrium properties of the models, see Ref.~\cite{Dimmelmeier:2005zk}.
  
  The fluid modes of this sequence were previously investigated in 
  different works, using various approaches~\cite{Font:2000rd,Stergioulas:2003ep,Dimmelmeier:2005zk}. 
  With our general-relativistic 3D simulations we are able to study the effect of
  rotation on the radial mode and compare the results with those obtained via
  approximated approaches. Our results are summarized in Table~\ref{tab:radialmodeAU}.
  We have found that the frequencies computed by Dimmelmeier et
  al.~\cite{Dimmelmeier:2005zk} in the conformally flat approximation 
  are consistent with ours (the difference is of the order of few
  percents); on the other hand, the results of Stergioulas et
  al.~\cite{Stergioulas:2003ep}, obtained in the Cowling approximation, 
  differ of about a factor two, consistently with the estimates of Ref.~\cite{1997MNRAS.289..117Y}.
  In all cases, the frequencies decrease if the rotation increases and 
  the trend is linear in the rotational parameter $\beta\equiv T/|W|$, 
  the ratio between the kinetic rotational energy and the gravitational potential energy.
  
  \begin{table}[t]
    \caption{\label{tab:radialmodeAU} 
      Frequencies of the fundamental radial mode of models in the sequence AU
      of uniformly rotating polytropic stars of Refs.~\cite{Stergioulas:2003ep} and
      \cite{Dimmelmeier:2005zk}.
      The frequency of model AU0 (A0) has also been computed in Ref.~\cite{Font:2001ew}
      ($1450$ Hz) and in this work ($1462$ Hz), where a finer grid was used (see Table~\ref{tab:radialmode}). 
      The data in the column marked as ``CF'' refer to Table~III of Ref.~\cite{Dimmelmeier:2005zk}.
      The data in the column marked as ``Cowling'' refer to Table~II of Ref.~\cite{Stergioulas:2003ep}.}
    \begin{ruledtabular}
      \begin{tabular}{lccc}
	MODEL & F [Hz] & F(CF) [Hz] & F(Cowling) [Hz]\\
	\hline
	\hline
	AU0 & 1444 & 1458 & 2706 \\
	AU1 & 1369 & 1398 & 2526 \\
	AU2 & 1329 & 1345 & 2403 \\
	AU3 & 1265 & 1283 & 2277 \\
	AU4 & 1166 & 1196 & 2141 \\
	AU5 & 1093 & 1107 & 1960 \\
      \end{tabular}
    \end{ruledtabular}
  \end{table}
  
  \subsection{Nonradial oscillations: comparing 1D and 3D metric waveforms}
  \label{sbsc:waveforms}
  
  Let us now turn to the discussion of nonspherical oscillations 
  and to the related extraction of waveforms from 1D and 3D simulations. 
  We consider a star perturbed with an $\l=2$, $H_{\l0}(r)$ profile with 
  $n=0$, according to the procedure outlined in Sec.~\ref{sbsc:init}.
  This system is evolved separately with the two codes and the
  related gravitational waveforms are compared.

\begin{figure}[t]
    \begin{center}
      \includegraphics[width=85 mm]{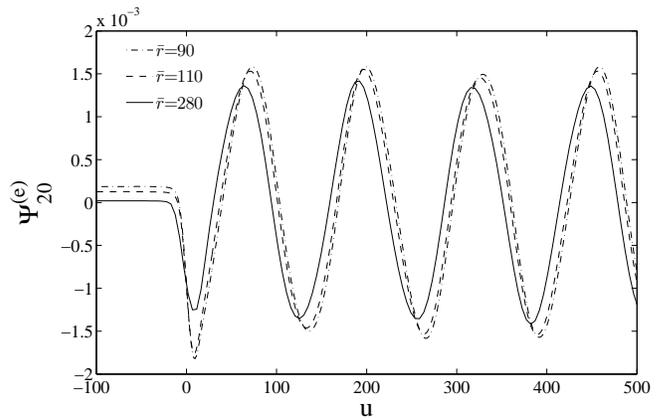}
      \caption{\label{fig:Psie_Pert_u} 
	Evolution of the Zerilli-Moncrief function,
	extracted at various isotropic radii $\bar{r}$, 
	versus the retarded time $u$, for the 1D linear code 
	evolution with $\lambda=\lambda1$.
	Note how the initial offset decreases with the extraction radius.}
    \end{center}
  \end{figure}

  We focus first on the discussion of the outcome of the 1D linear code.
  We accurately performed very long simulations, whose final time is 
  about $1$~s. The extraction radii for the Zerilli-Moncrief function extend as far as
  $\bar{r}=420$ ($\simeq 300M$). The resolution of the radial grid 
  is $\Delta r=0.032$, which corresponds to having 300 points inside the star.
  Fig.~\ref{fig:Psie_Pert_u} shows the Zerilli-Moncrief 
  function $\Psi^{(\rm e)}_{20}$ (for $\lambda=\lambda1$) extracted at different
  radii. It is plotted versus the observer retarded time, namely $u=t-r_*$,
  where $r_*$ is the Regge-Wheeler tortoise coordinate
  $r_*=r+2M\log[r/(2M)-1]$ and $M$ is the mass of the star.

  The farther observers that are shown in Fig.~\ref{fig:Psie_Pert_u} are 
  sufficiently deep in the wave-zone that the initial offset, that is 
  typically present due to the initial profile of $k_{20}$,  is small enough to 
  be considered negligible. We checked the convergence of the waves with the 
  extraction radius using as a reference point the maximum of 
  $\Psi^{(\rm e)}_{20}$. This point can be accurately fitted, as a function 
  of the extraction radius, with 
  \begin{equation}
    \max{\left(\Psi^{(\rm e)}_{20}\right)}\sim a^{\infty}+\frac{a^1}{r}\, .
  \end{equation}
  The extrapolated quantity $a^\infty$ allows an estimate of the error
  related to the extraction at finite distance
  \begin{equation}
  \delta a \equiv \frac{\Big |a^{\infty}-\max{\left(\Psi^{(\rm e)}_{20}(r)\right)}\Big
    |}{a^{\infty}} \, .
  \end{equation}
  The values of $\delta a$ for different radii are $\delta a\simeq0.5$ for $r=25M$, $\delta
  a\simeq0.09$ for $r=50M$, $\delta a\simeq0.017$ for $r=100M$ and $\delta a<0.016$ for $r>200M$.
  
  The waveform can be described by two different phases: (i) an initial transient, 
  of about half a gravitational-wave cycle, say up to $u\simeq 50$, related to the setup of
  the initial data
  \footnote{In practice, the first half cycle of the waves 
    cannot be expressed as a superposition of quasinormal modes and it is related 
    to the initial data setup. This initial transient is related to two facts: 
    (i) We use the conformally flat approximation;
    (ii) We assume $\dot{k}_\lm=0$ even if our
    initial configuration (a star plus a nonstatic perturbation) is evidently 
    not time symmetric, since a velocity perturbation is present and thus also a 
    radiative field related to the past evolutionary history of the star.},
  followed by (ii) a quasiharmonic oscillatory phase, where the matter
  dynamics are described in terms of the stellar quasinormal modes.
  From the Fourier spectrum of $\Psi^{(\rm e)}_{20}$ over a time
  interval from $1$ to about $30$ ms (namely $u\in[50,\,6000]$),
  we found that the signal is dominated 
  by the  $f$ mode (at frequency $\nu_f=1581$ Hz) with a much lower
  contribution of the first $p$ mode (at frequency around $\nu_{p_1}=3724$ Hz).
  The frequency of the $f$ mode agrees with that of Ref.~\cite{Font:2001ew}
  within $1-2$\%. The accuracy of our linear code 
  for frequencies obtained from Fourier analysis on such long time series has been 
  checked in Refs.~\cite{Nagar:2004av,Bernuzzi:2008fu} and is better 
  than 1\% on average.
  We mention that the Fourier analysis of the matter variable $H_\lm$
  permits to capture some higher overtones than the $p_1$ mode, although 
  they are essentially not visible in the gravitational-wave spectrum. 
  In a first approximation, the waveform can thus be thought
  as the superposition of damped harmonic oscillators
  \begin{equation}
    \label{eq:wavesTemplate}
    \Psi_{20} \sim \sum_{k=0}^{N} A_{2k}\cos(2\pi\nu_{2k}u+\phi_{2k})\exp(-\alpha_{2k}u) \ ,
  \end{equation}
  and we aim at determining the quantities $A_{2k}$, $\phi_{2k}$, $\nu_{2k}$ 
  and $\alpha_{2k}$ from a standard nonlinear least-square fit.  
  Since the frequency $\nu$ is also independently known from 
  the Fourier analysis, it is used as feedback for the fit.
  In addition, to quantify the global differences between
  the ``actual'' and the ``fitted'' time series, we compute
  the ($l^2$) scalar product
  \begin{equation}
    \label{eq:norm}
    \Theta(X,Y) = \frac{\sum_j 
      X_jY_j}{\sqrt{\sum_j \left(X_j\right)^2}
      \sqrt{ \sum_j \left(Y_j\right)^2}} \ ,
  \end{equation}
  which is bounded in the interval $[0,\,1]$, and then we look at the 
  residual ${\cal R}=1-\Theta$. This residual gives us a relative measure 
  of the reliability of the fit. In addition we use the $l^\infty$ distance:
  \begin{equation}
    {\cal D}(X,Y)=\max_j{\left|X_j-Y_j\right|}\,,
  \end{equation}
  that gives the maximum difference between the two time series.
  We will use the quantities ${\cal R}$  and ${\cal D}$
  also as measures of the global agreement between the 3D and 1D waveforms.
  \begin{table}[t]
    \caption{\label{tab:fitRes} Perturbative 1D evolution with $\lambda=\lambda1$:
      results of the fit to a superposition of two fluid modes
      [see Eq.~\eqref{eq:wavesTemplate}] over the interval $u\in[0,200000]$, \ie 
      about 1~s (\cf Fig.~\ref{fig:Psie_Pert_u2}). For this fit,
      ${\cal R}\simeq1.6\times10^{-6}$ and ${\cal D}\simeq3.6\times10^{-5}$ (see text for explanations).}
    \begin{ruledtabular}
      \begin{tabular}{lll}
         [arbitrary units] & 
         $\nu$ [Hz]   &
         $\alpha$ [$\rm s^{-1}]$ \\
	\hline
	\hline
	$A_{20}      $=$ 1.3145_{-2}^{+2}\times10^{-3}$ &
	$\nu_{20}    $=$ 1583.7369_{-1}^{+2}$ &
	$\alpha_{20} $=$ 3.7358_{-9}^{+9}$    \\
	$A_{21}      $=$ 3.517_{-12}^{+13}\times10^{-5}$ &
	$\nu_{21}    $=$ 3706.9413_{-11}^{+11}$ &
	$\alpha_{21} $=$ 0.421_{-6}^{+8}$       \\
      \end{tabular}
    \end{ruledtabular}
  \end{table}
  \begin{figure}[t]
    \begin{center}
      \includegraphics[width=85 mm]{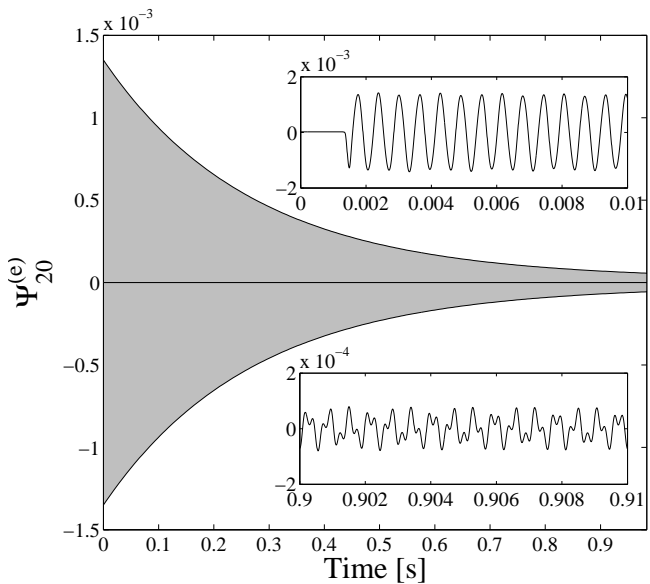}
      \caption{\label{fig:Psie_Pert_u2} 
    Evolution of the Zerilli-Moncrief function,
	extracted at isotropic radius $\bar{r}=280$ for the same evolution
	of Fig.\ \ref{fig:Psie_Pert_u} versus the retarded time $u$ expressed
	now in seconds (1 dimensionless time unit is equal to $4.92549\times10^{-6}$ s). 
        The main panel shows only the envelope of the waveform
	that is dominated by the damping time of the $f$ mode 
	($\tau_f\simeq0.27$~s). The two insets represent the full waveform at
	early (top) and late (bottom) times. The presence of the overtone is 
	evident in the oscillations at late times.}
    \end{center}
  \end{figure}
  
  On the interval $u \in[50,6000]$, the waves can be perfectly 
  (${\cal R}\simeq 7\times10^{-4}$, ${\cal D}\simeq 6\times10^{-6}$) represented by a one-mode expansion, $N=1$,
  as the waveform is dominated by $f$-mode oscillations.
  The frequency we obtain, $\nu_{20}=1580.79\pm0.01$ Hz, is perfectly
  consistent with that obtained via Fourier analysis; 
  for the damping time, we estimate  
  $\alpha_{20}=3.984\pm0.066$ $\rm s^{-1}$ and thus 
  $\tau_{20}=\alpha_{20}^{-1}\simeq0.25$ s. 
  If we consider the entire duration (1s) of the signal
  (see the inset in Fig.~\ref{fig:Psie_Pert_u2}), 
  it is clear that a one-mode expansion is not sufficient to accurately 
  reproduce the waveform. The Fourier analysis of the waveform in two 
  different time intervals, one for $t\lesssim0.5$ s and one for
  $t\gtrsim0.5$, reveals that in the second part of the signal the $p_1$ mode, 
  which has longer damping time, clearly emerges and must be taken into account.
  We fit the entire signal with two modes, namely $N=2$, with a global agreement 
  of ${\cal R}\simeq2\times10^{-6}$ and ${\cal D}\simeq4\times10^{-5}$.
  The results of the fit are reported in Table~\ref{tab:fitRes}. 
  The frequencies are slightly larger than those computed 
  via Fourier analysis and via the fit procedure restricted to only one mode on
  a shorter interval. They are, however, still consistent. The damping times are 
  $\tau_{20}=0.268$~s and $\tau_{21}=2.37$ s, with errors of 
  the order of $0.1$\% and $2$\% respectively.

  At this stage, we have clearly assessed the accuracy of the waveforms computed
  via our 1D code; in the following we shall consider these waveforms
  (extracted at the farthest observer) as {\it exact} for all practical purposes.
  We turn now to the discussion of the metric waveforms extracted from
  the 3D code and we compare them to the exact, perturbative
  results for different values of the perturbation $\lambda$.
  \begin{figure}[t]
    \begin{center}
      \includegraphics[width=85 mm]{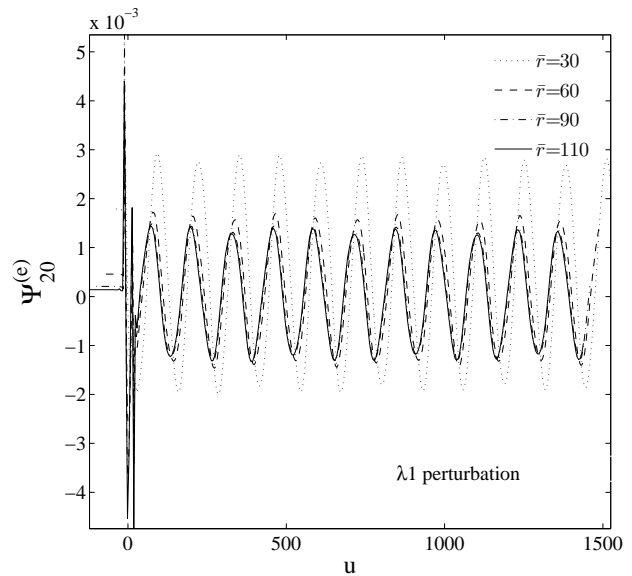}
      \caption{\label{fig:Psie_WE_u} 
        Zerilli-Moncrief normalized metric waveforms shown versus the
        observer retarded time $u=t-r_*$ at different extraction radii ($\bar{r}=30$ to
        $\bar{r}=110$), for a 3D evolution with perturbation $\lambda=\lambda1$.
        The initial part of the waveform is dominated by a pulse of 
        {\it junk} radiation.}
    \end{center}
  \end{figure}

  \begin{figure*}[t]
    \begin{center}
      \includegraphics[width=85 mm]{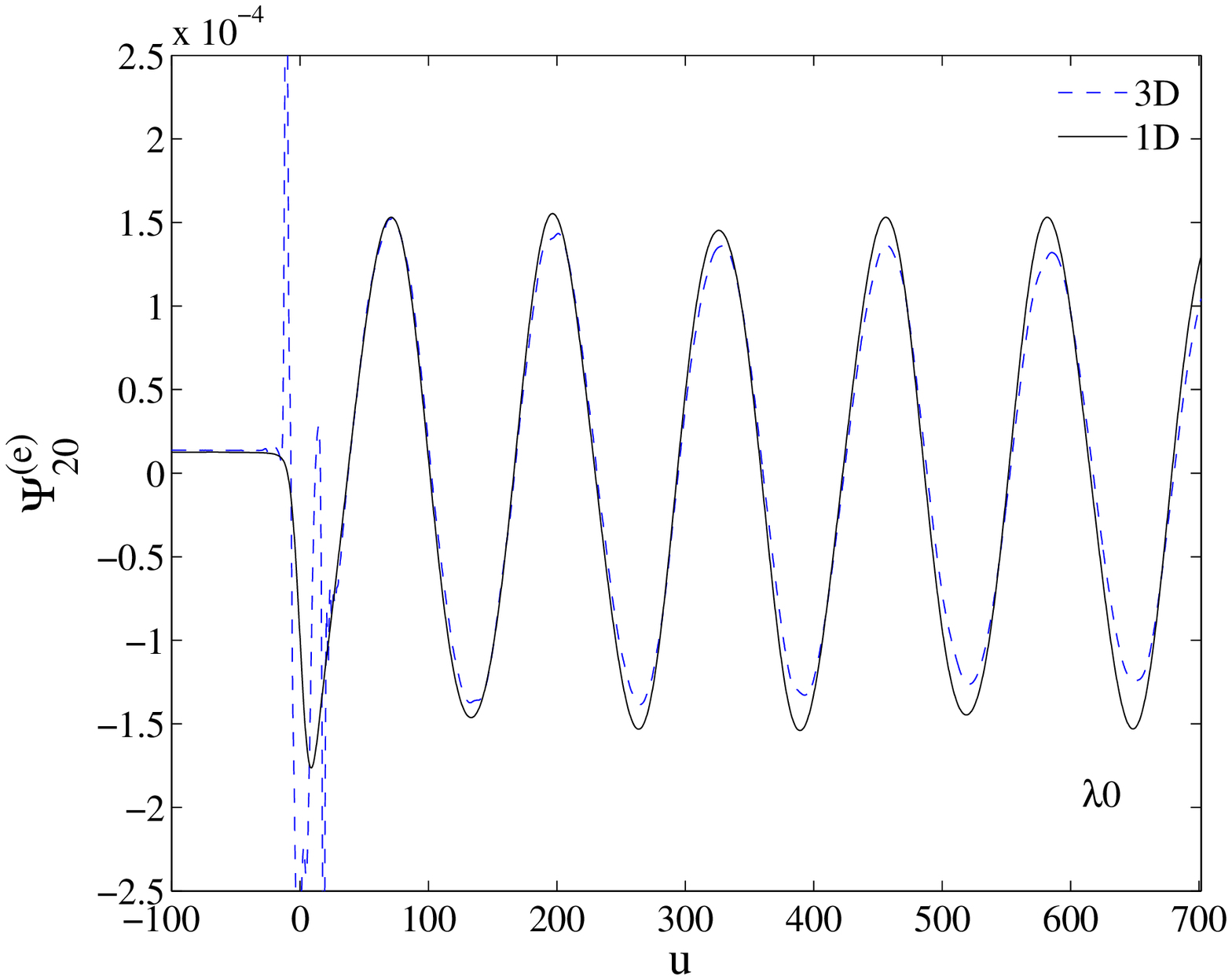}\hspace{1 mm} 
      \includegraphics[width=85 mm]{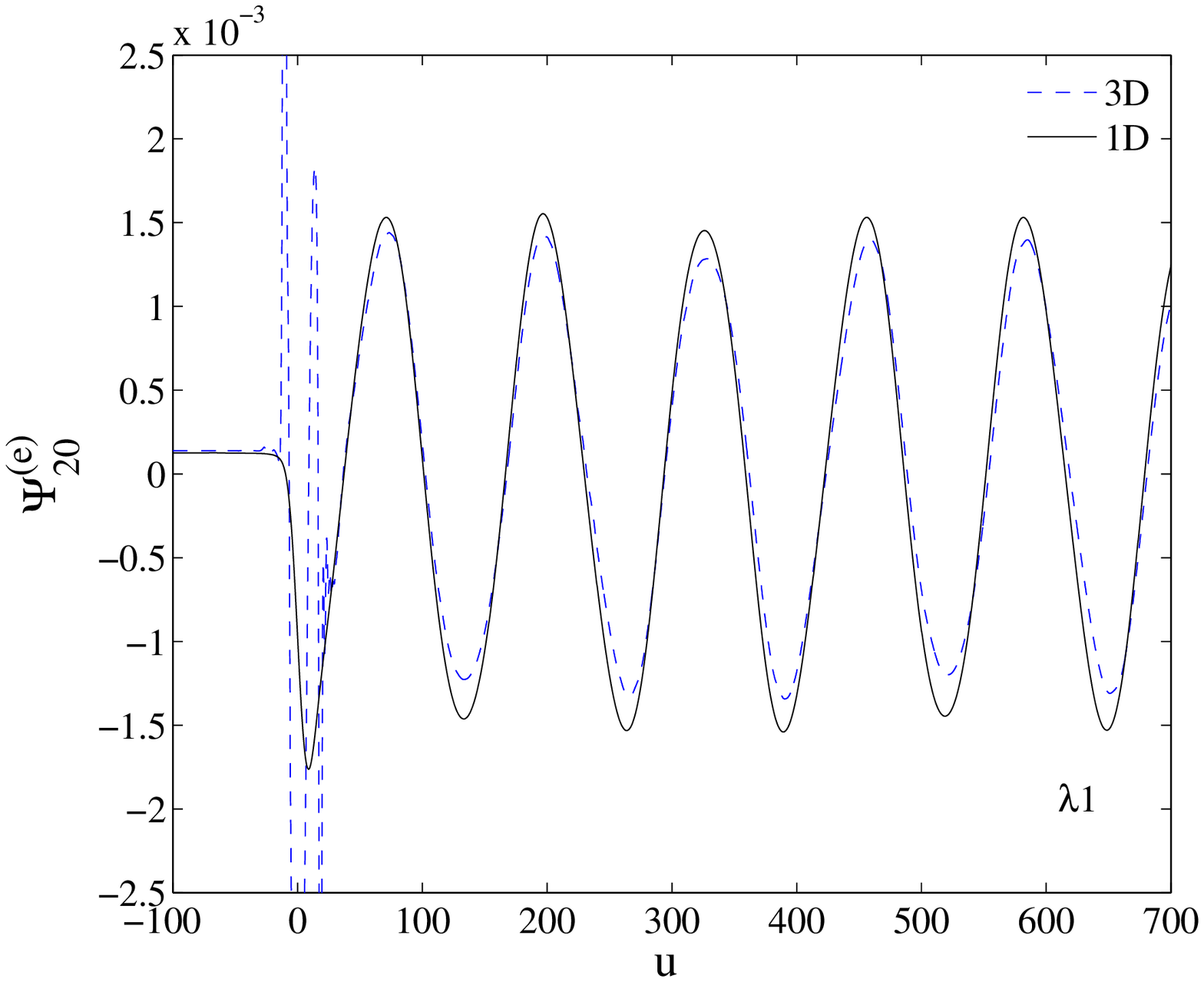}\\
      \includegraphics[width=85 mm]{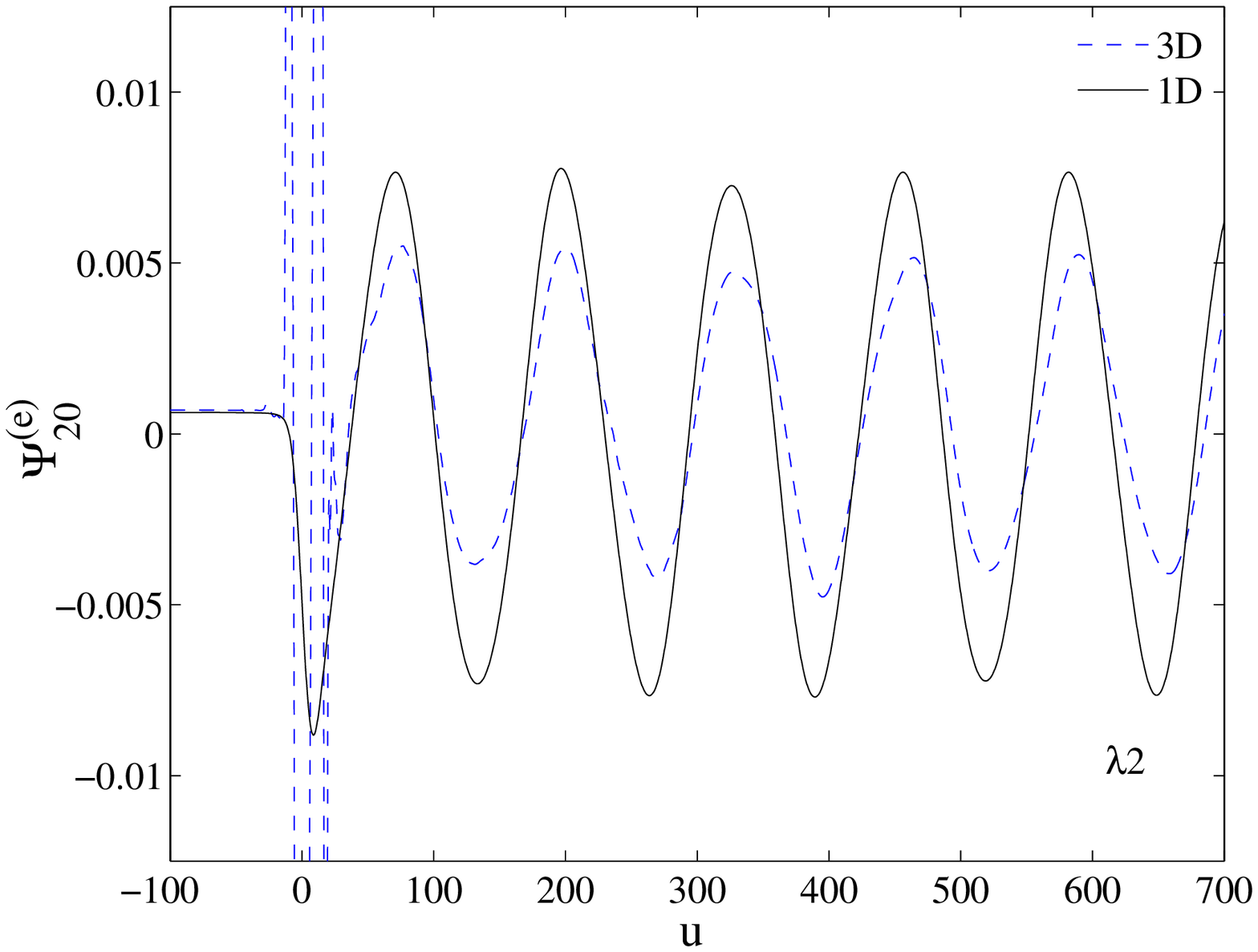}\hspace{1 mm} 
      \includegraphics[width=85 mm]{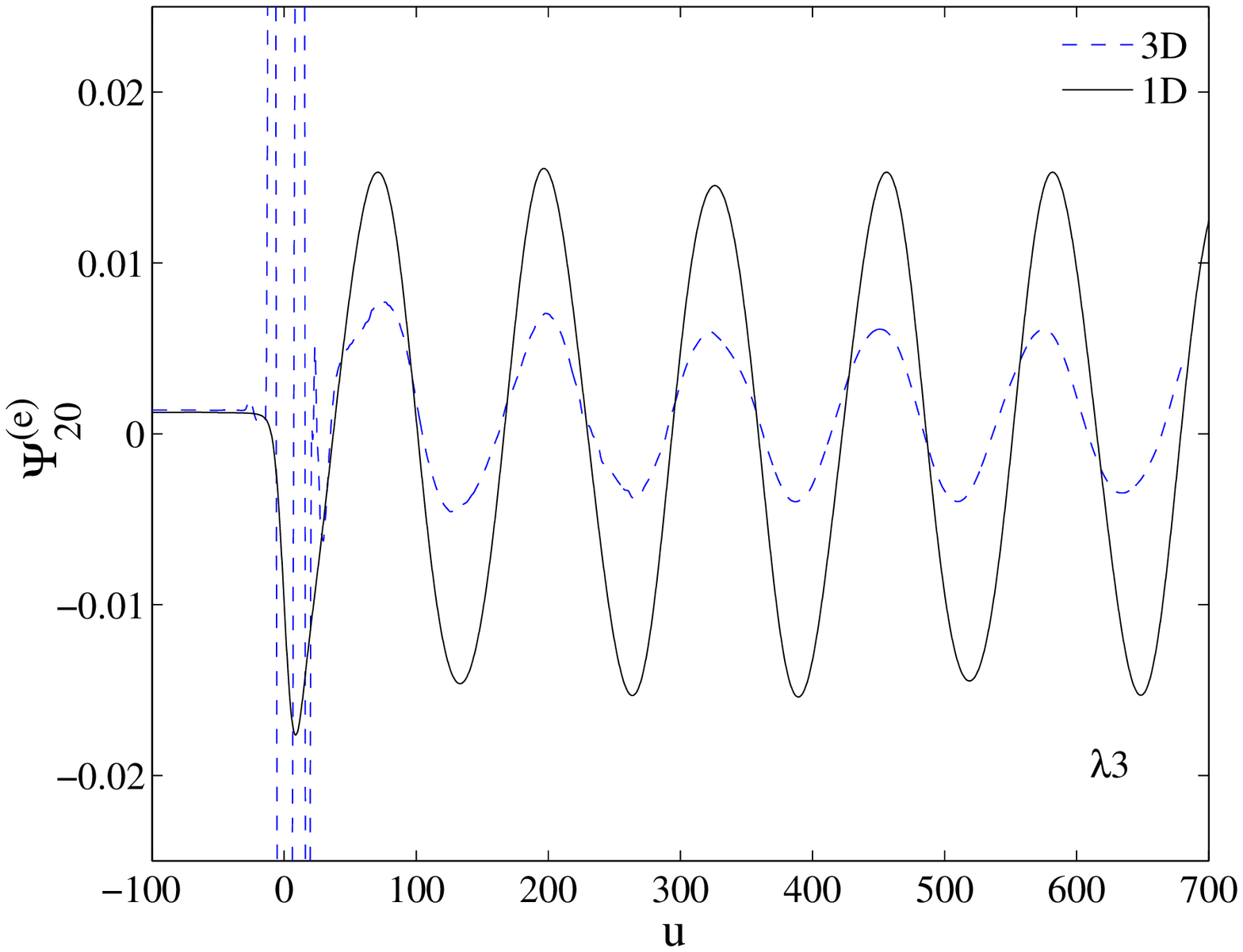}\\
      \caption{\label{fig:comp_Psi_u} (color online)
        Metric waveforms extracted at $\bar{r}=110$ computed
	from 3D simulations (dashed lines) and 1D linear simulations (solid lines) 
	for different values of the perturbation $\lambda$. If $\lambda$ is
	sufficiently small (\eg, $\lambda\lesssim 0.01$) the outcomes of the two codes 
	show good agreement. If the ``perturbation'' is large ($\lambda\sim0.1$),
	nonlinear effects become dominating.
      }
    \end{center}
  \end{figure*}
  The 3D simulations are performed over grids with
  three refinement levels and cubic boxes with limits 
  $[-120,\,120]$, $[-24,\,24]$ and $[-12,\,12]$ in each direction. The resolutions 
  of each box are $\Delta_{xyz}=0.5$, $0.25$ and $0.125$, respectively.
  Equations are evolved only on the first octant of the grid and symmetry conditions are applied.
  The outermost detector is located at isotropic-coordinate 
  radius $\bar{r}=110$ ($\sim80M$). 

  Figure~\ref{fig:Psie_WE_u} is obtained with perturbation $\lambda=\lambda1$.
  It displays the Zerilli-Moncrief normalized  metric waveforms, 
  extracted on coordinate spheres of radii $\bar{r}\in\{30,~60,~90,~110\}$
  and plotted versus the (approximate) retarded time $u = t-r_*$,
  where $r_* = r + 2 M \log[r/(2M)-1]$. Here, $r$ is the areal
  radius of the spheres of coordinate radius $\bar{r}$ and $M$
  is the Schwarzschild mass enclosed in 
  $\bar{r}$~\cite{Abrahams:1995gn,Allen:1998rg,Pazos:2006kz}.
  This figure is the 3D analogous of 
  Fig.~\ref{fig:Psie_Pert_u}. The 1D and 3D waveforms look 
  qualitatively very similar apart from the presence of 
  a highly-damped, high-frequency oscillation at early times.
  In Sec.~\ref{junk_RWZ} we will argue that 
  this oscillation is essentially unphysical because its amplitude 
  grows linearly with the extraction radius $\bar{r}$, 
  instead of approaching an approximately constant value (as it happens instead for the 
  subsequent fluid-mode oscillations). 
  Section~\ref{junk_RWZ} is devoted to a thorough discussion of
  these issues; for the moment, we simply ignore this problem
  and focus our attention only on the part of the waveform
  dominated by fluid modes.
  
\begin{figure}[t]
  \begin{center}
    \includegraphics[width=85 mm]{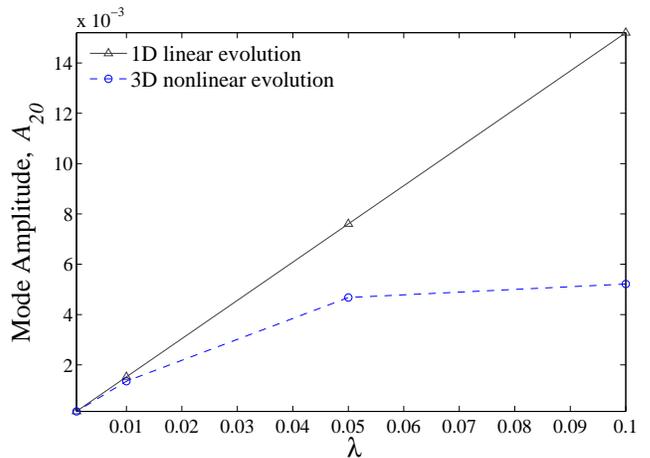}
    \caption{\label{fig:resAmp} (color online) Comparison between the oscillation amplitude 
      $A_{20}$ (from fits) in the 1D (linear) and 3D (nonlinear) simulations
      versus the initial perturbation $\lambda$.
      Deviations from linearity are occurring already for very small 
      values of $\lambda$.}
  \end{center}
  \end{figure}

  Each panel of Fig.~\ref{fig:comp_Psi_u} compares the 1D, exact 
  $\Psi^{(\rm e)}_{20}$ (dashed lines) with that computed via the
  3D code (solid lines)  for the four values of the perturbation $\lambda$.
  The extraction radius is (in both codes) $\bar{r}=110$
  and this implies that a nonzero, constant offset for $u\lesssim 0$
  is present. Note, in this respect, the good consistency between
  3D and 1D results for $u\lesssim 0$, confirming here the information
  enclosed in the bottom panel of Fig.~\ref{fig:ID:k_psie}.
  After the initial high-frequency (unphysical) oscillations, 
  the top-left panel of Fig.~\ref{fig:comp_Psi_u} shows that
  an excellent agreement between the waveforms is found when the
  perturbation is {\it small}. Then, for larger values of $\lambda$ 
  (until it assumes values that cannot be 
  considered a perturbation anymore) the amplitude of the oscillation 
  in the 3D simulations becomes smaller with respect to the 
  linear case, suggesting that nonlinear couplings 
  (specifically, couplings with overtones as well as couplings with the radial modes) 
  are redistributing the energy of the $\l=2$, $m=0$ 
  oscillations triggered by the initial perturbation. 
  In Sec.~\ref{sbsc:nnlin}, we will
  argue that couplings between modes become more and more relevant
  when the perturbation increases, giving a quantitative explanation
  to the phenomenology that we observe.
  This effect is summarized in 
  Fig.~\ref{fig:resAmp}, which displays the amplitude 
  $A_{20}$ obtained by fitting the waveform with 
  the template Eq.~\eqref{eq:wavesTemplate} versus the magnitude 
  of the perturbation for 1D (linear) and 3D (nonlinear) simulations.
  It is evident from the figure that there is a consistent deviation 
  from linearity already when the perturbation is relatively small 
  ($\lambda\lesssim 0.02$). As a measure of the global agreement 
  between 1D and 3D waveforms (as a function of the initial 
  perturbation $\lambda$) we list in
  Table~\ref{tab:globalagr} the $l^2$ residuals 
  ${\cal R}=1-\Theta(\Psi^{\rm 1D},\Psi^{\rm 3D})$ and 
  the $l^\infty$ distances ${\cal D}(\Psi^{\rm 1D},\Psi^{\rm 3D})$.
  \begin{table}[t]
    \caption{\label{tab:globalagr} ``Global-agreement'' measures computed 
      on the interval $\Delta u =[50,3000]$ (after the {\it junk} burst) at the
      outermost detector. Here, ${\cal R}=1-\Theta(\Psi^{\rm 1D},\Psi^{\rm 3D})$
      is the $l^2$ residual while  ${\cal D}(\Psi^{\rm 1D},\Psi^{\rm 3D})$ is
      the $l^\infty$ distance.}
    \begin{ruledtabular}
      \begin{tabular}{lcc}
	$\lambda$ & $\mathcal{R}$  & $\mathcal{D}$ \\
	\hline
	\hline
	$\lambda0$ & $3.07\times10^{-2}$ & $7.93\times10^{-5}$\\
	$\lambda1$ & $4.88\times10^{-2}$ & $7.79\times10^{-4}$\\
	$\lambda2$ & $1.63\times10^{-1}$ & $2.78\times10^{-3}$\\
	$\lambda3$ & $9.96\times10^{-1}$ & $2.04\times10^{-2}$
      \end{tabular}
    \end{ruledtabular}
  \end{table}

  \begin{table}[t]
    \caption{\label{tab:WEfreqs} Frequency analysis of the 3D waveforms
      (see Fig.~\ref{fig:comp_Psi_u}) over the interval $u\in[50,\,3000]$.
      The frequencies from 1D simulations are $\nu_f^{\rm 1D}=1581$ Hz 
      and $\nu_{p_1}^{\rm 1D}=3724$ Hz. From left to right the columns
      report the amplitude of the perturbation, the $f$-mode frequency, its relative
      difference with the 1D value, the $p_1$-mode frequency and
      its relative difference with the 1D value.}
    \begin{ruledtabular}
      \begin{tabular}{lcccc}
	$\lambda$ & $\nu_f^{\rm 3D}$ [Hz] & Diff.[\%] & $\nu_{p_1}^{\rm 3D}$ [Hz] & Diff.[\%]\\
	\hline
	\hline
	$\lambda0$ & 1578 & 0.2 & 3705 & 0.5\\
	$\lambda1$ & 1576 & 0.3 & 3705 & 0.5\\
	$\lambda2$ & 1573 & 0.5 & 3635 & 2.4\\
	$\lambda3$ & 1623 & 2.7 & 3565 & 4.3
      \end{tabular}
    \end{ruledtabular}
  \end{table}

  The 3D waveforms for $\lambda0$ and $\lambda1$ turn out to be damped 
  on a time scale of about 20~ms. This damping time is much shorter 
  than the one of the $f$ mode or $p_1$ mode, as computed via the 1D approach.
  This effective-viscosity damping time $\tau^{\rm visc}$ 
  (that is related to the inverse of the viscosity coefficient)
  can be extracted by means of the fit analysis discussed above for the 
  waveform. We have found that $\tau^{\rm visc}$
  depends on the initial perturbation, being $\tau^{\rm visc}\simeq 0.022, 0.132, 0.203, 0.129$~s 
  respectively for $\lambda=\lambda0,\lambda1,\lambda2,\lambda3$. 
  The best agreement with the expected physical value of $\tau_{20} = 0.268$~s
  is obtained for $\lambda=\lambda2$; both for larger and smaller perturbations the 3D results show
  even shorter damping times. The errors on these quantities 
  are of the order of $0.5$\%. The interpretation of these results may include two different
  effects. The smaller damping time of the wave for the $\lambda=\lambda3$ perturbation with respect
  to the  $\lambda=\lambda2$ one may be interpreted as due to the nonlinear
  couplings that allow the disexcitation of the fundamental mode in other
  channels; as it can be seen from Fig.~\ref{fig:resAmp} and
  Fig.~\ref{fig:couplings}, the importance of nonlinear effects is larger for the simulation with 
  perturbation $\lambda=\lambda3$.
  However, for perturbations smaller than $\lambda=\lambda2$ the effective viscosity is not found
  to decrease towards the expected perturbative value, as it could have been expected from the above
  argument. This discrepancy might be due to the numerical 
  viscosity proper of the evolution scheme. Such numerical viscosity would have a bigger influence in
  low-perturbation simulations, where the energy lost from the fundamental mode into other modes is
  smaller (while in higher-perturbation simulations the coupling of modes is the dominant effect). 
  Although the detailed analysis of the numerical viscosity of the 3D code 
  is beyond the scope of the present work, we checked that, as expected, 
  it depends on the grid resolution. We performed tests using a three-refinement-level 
  setup with the resolution of the coarsest grid (with limits $[-120,120]$ 
  in the three directions) set at the values $\Delta_{xyz}=2$ (\emph{low}), 
  $1$ (\emph{medium}) and $0.5$ (\emph{high}). Using these three resolutions, 
  we observed that, in the case of the coarsest grid, there was an initial explosion in the amplitude, 
  then followed by a strong damping during the first 
  five gravitational-wave cycles. This shows that this resolution is not even sufficient 
  to extract the qualitative behavior of the waveform.
  On the other hand, the other two resolutions did not show any qualitative difference 
  in addition to the different value of the ``effective viscosity'', that is smaller 
  for higher resolutions.
  We also checked whether there is a measurable effect due to the 
  artificial atmosphere. Focusing only on the $\lambda=\lambda0$ perturbation,
  we varied the value of the rest-mass density of the atmosphere in the range 
  $\rho_{\rm atm}=10^{[-5,-6,-7]}\rho_{\rm max}$, without finding any 
  significant influence on the values of $\tau^{\rm visc}$.
  We leave to forthcoming studies a detailed analysis of the viscosity of the 3D evolution code.

  Finally, we have also Fourier-transformed the 3D waveforms to extract
  the fluid-mode frequencies and we have compared them with the linear
  ones. This comparison is shown in Table~\ref{tab:WEfreqs}. 
  Apparently, the frequency of the $f$ mode
  (that dominates the signal) is less sensitive to nonlinear effects than
  its amplitude, as it can be seen from the fact that only the $\lambda=\lambda3$ initial data are such to force the 
  star to oscillate at a frequency slightly different from that of the 
  linear approximation. On the other hand, the first overtone (the $p_1$ mode)
  seems more sensitive. It is in any case remarkable that for $\lambda=\lambda0$ and $\lambda=\lambda1$ 
  the frequencies from 3D and 1D simulations coincide at better than 1\%,
  suggesting that the main gravitational-wave frequencies are only mildly
  affected by nonlinearities.

  \subsection{Nonradial oscillations: comparing 1D and 3D curvature waveforms}
  \label{sbsc:waveformsPsi4}

  This section is devoted to the comparison between 1D and 3D {\it curvature}
  waveforms. In the 1D code one can use the relation
  \begin{equation}
    \label{rpsi4}
    r\psi_4^{\lm}= r\ddot{h}^\lm = N_\l\left(\ddot{\Psi}_{\lm}^{(\rm e)} 
                    + \i \ddot{\Psi}_{\lm}^{(\rm o)}\right)
  \end{equation}
  to obtain the Newman-Penrose scalar (multiplied by the extraction radius) 
  $r\psi_4^\lm $ from the gauge-invariant metric master functions. 
  Because of our choice of initial conditions, we shall consider only 
  $\Psi^{(\rm e)}_{20}$ in the following.~
  \footnote{Note that in principle one could compute $\psi_4$
    independently, solving the Bardeen-Press-Teukolsky 
    equation \cite{Pons2002}.}
  The second time-derivative of $\Psi^{(\rm e)}_{20}$ is computed via
  finite differencing, by applying twice a first-order
  derivative operator with 4th-order accuracy.
  By contrast, in the 3D code $\psi_4^{20}$ is extracted {\it independently} 
  of the metric waveform. Then, one computes $r\psi_4^{20}$, where 
  $r$ is an approximated radius
  \footnote{This is an approximate relation as $\bar{r}$ is a
    coordinate radius and the mass inside the sphere of radius $\bar{r}$ 
    is time dependent. We neglect all higher-order effects here 
    as this approximation is sufficiently accurate for our purposes.}
  from Eq.~\eqref{r_from_barr} with $M=1.4$.
  
  \begin{figure}[t]
    \begin{center}
      \includegraphics[width=95 mm]{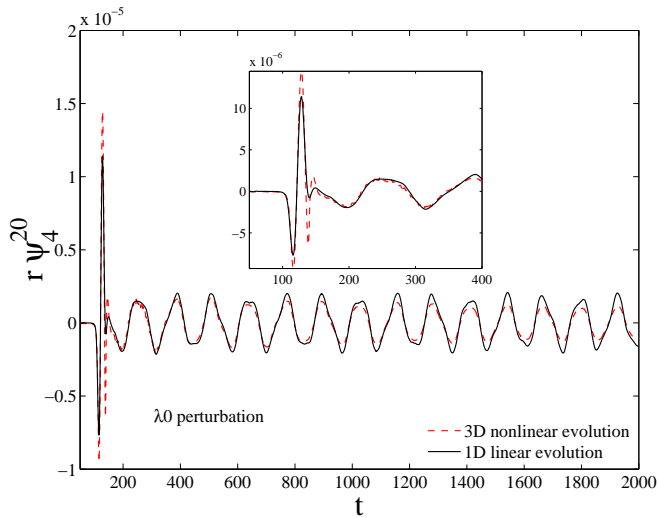}
      \caption{\label{fig:psi4} (color online) 1D versus 3D evolution of the
	$r\psi_4^{20}$ curvature waveform for perturbation
	$\lambda=\lambda0$. The initial transient is consistent between the two
        evolutions. The inset concentrates on the initial part of the waveform.}
    \end{center}
  \end{figure}
  
  Figure~\ref{fig:psi4} displays
  the $r\psi_4^{20}$ waveforms from 1D (solid line) and 3D (dashed line) 
  evolutions with perturbation  $\lambda=\lambda0$. 
  The extraction radius is $\bar{r}=110$ in both codes.
  Visual inspection of the figure immediately suggests that:
  (i) The initial transient in the 1D metric waveform preceding the setting in of the
  quasiharmonic $f$-mode oscillation results in a highly
  damped, high-frequency oscillation; (ii) The initial transient radiation has 
  {\it the same qualitative shape} in both the 1D and 3D waveforms,
  although the amplitude of the oscillation is larger in the latter 
  case.
  At this point one should note that: (i) In the 1D case, although
  the conformally flat condition is imposed at $t=0$, the constraint
  is solved {\it numerically} and thus a small violation of this
  condition occurs;
  (ii) The violation is expected to be 
  larger in the 3D case, because of the larger truncation errors.
  It is in any case remarkable that, as the figure shows, these errors
  (\eg the slightly different shapes of $k_{20}$, the linear interpolation  
  from spherical to Cartesian coordinates, etc.) are sufficiently  
  under control to produce the same qualitative behavior  
  besides {\it small quantitative differences} 
  in the initial part of the 1D and 3D waveforms. 
  \begin{figure}[t]
    \begin{center}
      \includegraphics[width=85 mm]{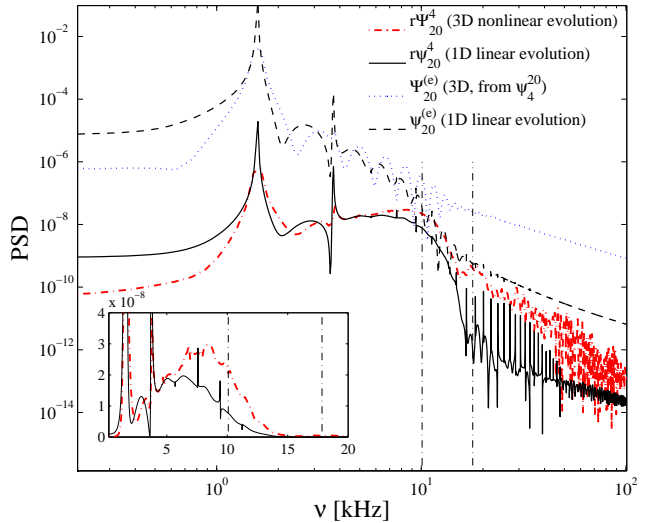}
      \caption{\label{fig:PSD_psi4} (color online)
	Comparison of the PSD of various waveforms.
	The PSD of the $r\psi_4^{20}$ and $\Psi^{(\rm e)}_{20}$ from the 1D code are 
	compared with those from the 3D code, obtained after integration. The initial 
	perturbation is $\lambda=\lambda0$. The $w$-mode frequencies $w_1$ and $w_2$ 
	are superposed to the spectra as vertical dot-dashed lines. See text for discussions.}
    \end{center}
  \end{figure}

  The question that occurs at this point is whether the violation of the conformally
  flat condition introduces some amount of physical $w$-mode excitation in the waveforms.
  To answer this question we show in Fig.~\ref{fig:PSD_psi4} the 
  Fourier power spectral density (PSD)
  of the $r\psi_4^{20}$ waveforms of Fig.~\ref{fig:psi4}.
  The PSD is computed all over the waveform and not
  only during the ``ring-down'', because of the difficulty of separating reliably
  this part from the ``precursor''~\cite{Bernuzzi:2008rq}.
  We are aware of the problems related to the precise 
  determination the $w$-mode frequencies and to their location in the waveform 
  (see Refs.~\cite{Bernuzzi:2008rq,1999CQGra..16R.159N} for a related discussion), 
  and in particular of the fact that the Fourier analysis can not provide 
  accurate and definitive answers, essentially because, in the presence of 
  damped signals with frequencies comparable to the inverse of the damping time, 
  the Fourier spectrum results in a broad peak.
  However it represents a fundamental part of the analysis 
  and, in the present case, is preferable to a fitting procedure 
  because of the already mentioned problem of separating the precursor from the ring-down
  part.
  
  The dashed-dotted vertical lines of Fig.~\ref{fig:PSD_psi4} locate the first two 
  $w$-mode frequencies of this model, $\nu_{w_1}=10.09$ kHz and
  $\nu_{w_2}=17.84$ kHz. These frequencies have been computed by
  K.~Kokkotas and N.~Stergioulas via an independent frequency-domain code 
  and have been kindly given to us for this specific comparison.
  Two of the maxima of the PSD of the 3D $r\psi_4^{20}$ waveform can be associated to 
  the frequencies $\nu_{w_1}$ and $\nu_{w_2}$, even if they are in a region 
  very close to the noise.
  The frequency $\nu_{w_1}$ is probably slightly excited also in 
  the 1D case (see inset), while only noise is present around $\nu_{w_2}$.
  In Fig.~\ref{fig:PSD_psi4} we show also the PSD of the 
  1D $\Psi^{(\rm e)}_{20}$ (dashed dark line) and the one of the 
  3D $\Psi^{(\rm e)}_{20}$ (dashed light line) obtained from the double time-integration 
  of $\psi_4^{20}$.
  In both cases, it is not possible to disentangle $\nu_{w_1}$ and 
  $\nu_{w_2}$ from the background noise.

  The fact that a signal characterized by highly damped modes is much less 
  evident in the PSD of the metric waveform than in the corresponding curvature one
  is simply due to the second derivative that relates the two gauge-invariant
  functions. When space-time modes are excited, the metric waveform is 
  (approximately) composed by a pure ring-down part plus a tail 
  contribution~\cite{Price:1971fb}, that is $\Psi^{(\rm e)}_{20}\approx e^{-\sigma t}+\beta t^{-7}$, 
  where $\sigma=\alpha+i\omega$ ($\alpha$ is the inverse of the 
  damping time and $\omega$ the $w$-mode frequency) and $\beta$ is a numerical
  coefficient. When one takes two time derivatives to
  compute $r\psi_4^{20}$ from $\Psi^{(\rm e)}_{20}$, the tail contribution
  is suppressed by a factor $t^{-2}$ and the oscillatory part 
  of the waveform emerges more sharply.
  This comparison suggests that the best
  way to extract information about $w$ modes (especially when their
  contribution is small) is, in general, to look at $r\psi_4^\lm$.
  In addition, it also highlights that, while it is not possible to exclude
  the presence of $w$ modes in the $r\psi_4^\lm$ signal due to the small violation 
  of the conformally flat condition at $t=0$, at the same time we can not 
  definitely demonstrate that those high frequencies present near the noise 
  are attributable to $w$ modes. 
  In the next section we are going to show similar analyses on 
  the spectra computed from $\Psi^{(\rm e)}_{20}$ waveforms 
  extracted \'a la Abrahams-Price from the 3D simulation.

  Finally, the global-agreement measures on the $\psi_4$ extraction 
  are $\mathcal{R}\simeq1.42\times10^{-2}$ and $\mathcal{D}\simeq9.09\times10^{-7}$,
  and they highlight some differences between the linear and the nonlinear
  approach. 

  The analysis discussed so far indicates that, in the present framework,
  the wave-extraction procedure based on the Newman-Penrose scalar $\psi_4$ 
  seems to produce waveforms that, especially at early times, are more accurate
  than the corresponding ones extracted via the Abrahams-Price metric-perturbation 
  approach. However, one of the big advantages of the latter
  method is that the waveforms $h_+$ and $h_\times$  
  are directly available at the end of the computation, and thus ready to be 
  injected in some gravitational-wave--data-analysis procedure.
  By contrast, if we prefer to use Newman-Penrose wave-extraction procedures
  (which are the most common tools employed in numerical-relativity simulations
  nowadays), we must consistently give prescriptions 
  to obtain $\Psi^{(\rm e/o)}_{\lm}$ from $\psi_4^\lm$.
  To do so, one needs to perform a double (numerical) time integration, 
  with at least two free integration constants to be determined 
  to correctly represent the physics of the system.
  Inverting Eq.~\eqref{rpsi4} following the considerations of 
  Sec.~\ref{subs:NPform}, we obtain the following result [see Eq.~(\ref{eq:psi4const})]:
   \begin{align}
    rh^\lm&=N_\l\left(\Psi^{({\rm e})}_{\lm}+\i\Psi^{(\o)}_{\lm}\right)\\
    &=\int_{0}^{t}dt'\int_{0}^{t'}dt'' r\psi^4_\lm(t'') + Q_0 + Q_1 t + Q_2 t^2,
   \nonumber \\
    &=r\tilde{h}_\lm(t) + Q_0 + Q_1 t + Q_2 t^2,
   \nonumber
  \end{align}
  where $Q_0$, $Q_1$ and $Q_2$ are (still) undetermined
  integration constants, which are complex if $m\neq 0$. Note that this relation does not involve
  the $Q_2$ integration constant only if finite-radius extraction effects can be considered
  negligible (see below).

  \begin{figure}[t]
  \begin{center}
   \includegraphics[width=85 mm]{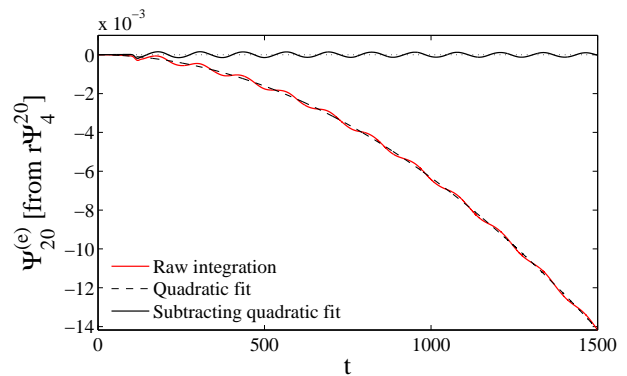}  
   \caption{\label{fig:psie_from_psi4_bare} (color online) Recovery of 
   $\Psi^{(\rm e)}_{20}$ from two successive time integrations of
   $r\psi_4^{20}$ extracted from the 3D simulation with initial perturbation $\lambda=\lambda0$.
   After the subtraction of a quadratic floor the waveform 
   correctly oscillates around zero.}
    \end{center}
  \end{figure}

  Our aim is to recover the metric waveform that corresponds to
  the 3D $r\psi_4^{20}$ waveform that we have characterized above.
  We consider the waveform of Fig.~\ref{fig:psi4} up to $t=1500$,
  where the reduction in amplitude due to numerical viscosity is
  already of the order of $30\%$ with respect to the exact
  linear waveform.
  This sampled curvature waveform is integrated twice in time, 
  from $t=0$ without fixing any integration constant to obtain
  $r\tilde{h}_\lm(t) $.

  \begin{figure}[t]
    \begin{center}
      \includegraphics[width=85 mm]{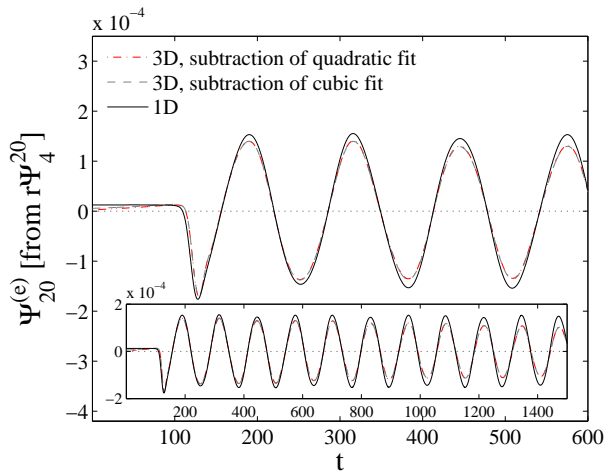}
    \end{center}
    \vspace{-5mm}
      \caption{\label{fig:psie_from_psi4} (color online) 
	Selecting the best fitting function for the 3D integrated-and-subtracted metric 
	waveform. The dash-dotted line refers to the curve obtained with the subtraction of a cubic fit
        and the dashed line to the curve obtained with the subtraction of a quadratic fit. They are
        very similar to each other and to the 1D waveform (solid line). 
        Contrary to the case of the extracted metric waveform,
	no initial burst of radiation is present in the
	3D integrated-and-subtracted waveform.
        } 
  \end{figure}

  The raw result of this double integration is shown in 
  Fig.~\ref{fig:psie_from_psi4_bare}. The 
  ``average'' of the oscillation does not lay on a straight 
  line, as it does instead in the case of the waveforms of binary black-hole
  coalescence discussed in Ref.~\cite{Damour:2008te},
  but rather it shows also a quadratic correction due to the finite 
  extraction radius (see discussion in Sec.~\ref{subs:NPform}).
  
  Indeed, when a ``floor'' of the form $P(t)=Q_0 + Q_1 t + Q_2 t^2$ is subtracted,
  the resulting metric waveform is found to oscillate around zero, as it can be seen in
  Fig.~\ref{fig:psie_from_psi4_bare} and Fig.~\ref{fig:psie_from_psi4},
  which focuses on the beginning of the oscillation.
  The values of the coefficients of $P(t)$ obtained from the fit 
  are $Q_0 =-4.338\times 10^{-7} $, $Q_1= -1.2462\times 10^{-7}$ and $Q_2=-6.2046\times
  10^{-9}$. The fact that $Q_0<0$ is connected to the choice of initial data we
  made (\ie $k_{20}\neq 0$ at $t=0$). 
  Then, $Q_1\neq 0$ indicates that the system is (slightly) out of equilibrium 
  already at $t=0$ and it is thus emitting gravitational waves since
  $\dot{\Psi}^{(\rm e)}_{20}(0)\neq 0$. This is consistent with the choice of 
  initial data we made, that is a perturbation that appears instantaneously 
  at $t=0$ without any radiative field obtained from the solution of the
  momentum constraint (since we use time-symmetric initial perturbations, for which
  $\dot{k}_{20}=\dot{\chi}_{20}=0$). 
    
  We tested the robustness of the quadratic fit
  by adding a cubic term $Q_3 t^3$ to $P(t)$
  and then fitting again. In Fig.~\ref{fig:psie_from_psi4},
  we compare the 3D $\Psi^{(\rm e)}_{20}$ waveform
  corrected with a cubic fit (dashed line) with the one corrected
  with a quadratic fit (dash-dotted line) and with the
  ``exact'' 1D metric waveform (solid line) output by the {\tt PerBACCo} code.
  Note that the 1D waveform has been suitably timeshifted
  in order to be visually in phase with the others at the beginning of the
  simulation. The figure suggests that the effect of the cubic 
  correction is almost negligible (one only finds slight
  changes in the very early part of the waveform). 
  The values of the fitting coefficients $Q_i$ are
  $Q_0 = -1.0096\times 10^{-5}$, $Q_1=-6.3331\times 10^{-9}$,
  $Q_2=-4.747\times 10^{-8}$, $Q_3=6.7114\times 10^{-14}$.
  The fact that $Q_3$ is many orders of magnitude smaller than the other coefficients is a good indication 
  that the quadratic behavior is indeed the best choice here.
  Consistently with the curvature waveform of Fig.~\ref{fig:psi4},
  we note the excellent agreement between 1D and 3D (integrated) metric
  waveforms {\it also} in the initial part of the waveform, \ie up to 
  $t\simeq 200$ (corresponding to the high-frequency oscillation
  in $r\psi_4^{20}$). Evidently, this is in contrast with the 
  Abrahams-Price metric waveform in the top-left panel of
  Fig.~\ref{fig:comp_Psi_u} (we will elaborate more on this in the
  next section). 

  Finally, we point out that the coefficient $Q_2(r)$ shows, 
  as expected, a clear trend towards zero for increasing values of the extraction radius.

  \subsection{Advantages and disadvantages of the Abrahams-Price metric 
              wave-extraction procedure}
  \label{junk_RWZ}
    
\begin{figure}[t]
  \begin{center}
    \includegraphics[width=85 mm]{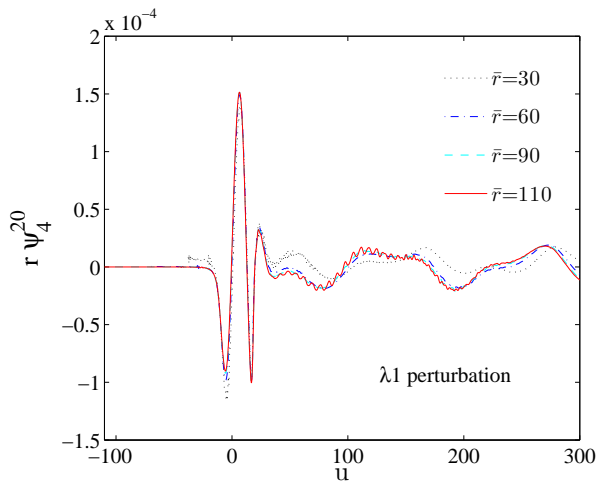} 
  \end{center}
  \vspace{-5mm}
    \caption{\label{fig:psi4_peeling} (color online) The quantities 
      $r\psi_4^{20}$ extracted at different radii are superposable 
      (as expected from the ``peeling'' theorem). See text for further explanation.}
  \end{figure}

\begin{figure}[t]
  \begin{center}
    \includegraphics[width=85 mm]{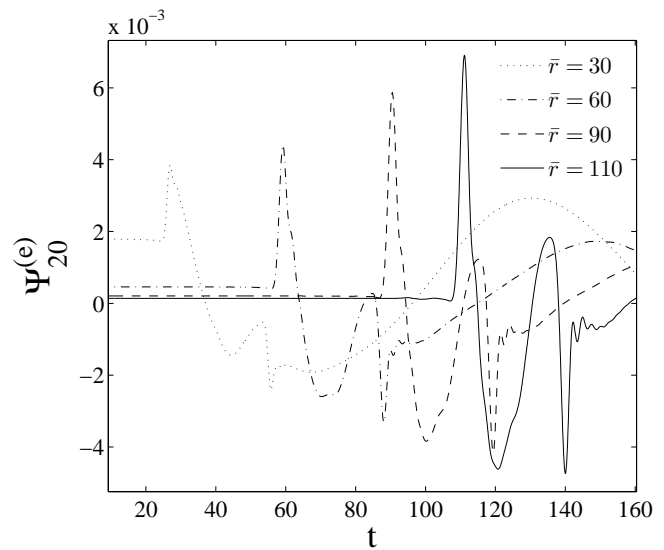}
    \caption{\label{fig:Psie_WE_u-JUNK} Metric waveforms computed via the Abrahams-Price procedure
      and extracted at different radii, for a 3D evolution with perturbation
      $\lambda=\lambda0$. Note how the amplitude of the initial burst grows linearly with $\bar{r}$.
  }
  \end{center}
  \end{figure}

  The analysis carried out so far suggests that both
  the Regge-Wheeler-Zerilli metric-based and the Newman-Penrose
  $\psi_4$-curvature-based wave-extraction techniques can be employed
  to extract reliable gravitational waveforms from simulations of compact 
  self-gravitating systems.
  For the particular case of an oscillating neutron star as considered
  here, both extraction methods allow to obtain 
  waveforms that are in very good agreement with the linear
  results. Despite this success, the two approaches are not free 
  from drawbacks.
  Let us first focus on $r\psi_4^\lm$ curvature waveforms.
  The comparison between 1D and 3D $r\psi_4^{20}$ waveforms 
  in Fig.~\ref{fig:psi4} (as well as between integrated 
  metric waveforms in Fig.~\ref{fig:psie_from_psi4}) shows good
  consistency between the two (as long as the effects of numerical viscosity
  on the evolution of the system remain negligible). 
  As we mentioned above, we think that the most 
  important information enclosed in Fig.~\ref{fig:psi4} is that
  the differences between the high-frequency oscillations
  in the initial part of the waveforms (where $w$ modes 
  are probably present in the 3D case) are {\it small}. 
  This fact makes us confident that the violation of the 3D 
  Hamiltonian constraint at $t=0$ (due to its approximate 
  solution\footnote{We recall that the 3D Hamiltonian 
    constraint is solved at the linearized level on an isotropic
    grid and then the resulting metric perturbation is interpolated 
    on the Cartesian grid. Typically, this procedure leads to
    larger errors than if solving the constraints directly on the
    Cartesian grid.})
  as well as the violation of the conformally flat condition
  are {\it sufficiently small} to avoid pathological behavior during
  evolution.
  A further confirmation of the accuracy of the evolution and of the
  curvature extraction is given by Fig.~\ref{fig:psi4_peeling}: 
  The quantities $r\psi_4^{20}$ extracted at various radii
  ($\bar{r}\in\{30,60,90,110\}$) and plotted versus retarded time
  are all superposed. This confirms the theoretical 
  expectations of the peeling theorem~\cite{Stewart:1991} and indicates (once more) 
  that the quantity $r\psi_4^{20}$ is accurately computed.
  In Fig.~\ref{fig:psi4_peeling}, $r$ is obtained from $\bar{r}$ via
  Eq.~\eqref{r_from_barr}. The retarded time is approximated with
  the standard $r_*$, where the constant mass $M=1.4$ has been used.
  In our setup, the only subtle issue about $r\psi_4^{\lm}$
  seems to be the computation of the corresponding metric
  waveform via a double time integration. Although we were able to 
  obtain a rather accurate metric waveform, the time-integration 
  procedure (including the evaluation of the integration constants) may not be 
  likewise straightforward in other physical settings.
  By contrast, the Abrahams-Price wave-extraction procedure
  directly produces the metric waveform and no time integrations are
  needed. For this reason, it looks {\it a priori} more appealing
  than $\psi_4$ extraction.
  Unfortunately, the results that we have presented so far 
  (notably our Fig.~\ref{fig:comp_Psi_u}) indicate
  that this computation can be very delicate and can give 
  unphysical results even in a very simple system like an 
  oscillating polytropic star: we have found that $\Psi^{(\rm e)}_{20}$
  extracted in this way is unreliable at early
  times, because of the presence of high-frequency, highly damped 
  oscillations, that are instead absent in both the 1D
  linear metric waveforms and the 3D metric waveforms time-integrated 
  from $r\psi_4^{20}$. 
  The unphysicalness of this initial ``burst'' of radiation is 
  evident from Fig.~\ref{fig:Psie_WE_u-JUNK}, where the extractions at various radii
  $\bar{r}\in\{30,60,90,110\}$ of the quantity 
  $\Psi^{(\rm e)}_{20}$ are compared: the amplitude {\it grows} with $\bar{r}$, instead of decreasing 
  progressively to approach a constant value (as it is the case for the
  $f$-mode--dominated subsequent part of the waveform)
  \footnote{To assess this statement we have also 
  performed simulations with extraction radii up to $\bar{r}=200$.}.
  The weird behavior at early times of the extracted $\Psi^{(\rm e)}_{20}$ 
  indicates that this function does not satisfy 
  the Zerilli equation in vacuum. Consistently, the perturbative
  Hamiltonian constraint in vacuum, Eq.~\eqref{iso_constraint} with $H_{20}=0$, 
  constructed from the 3D metric multipoles $(\chi_{20},k_{20})$,  
  must be violated of some amount in correspondence of the
  {\it junk}
  \footnote{The Zerilli equation, and thus the
    Zerilli-Moncrief master function, is obtained by combining together the
    perturbative Einstein equations, one of which is precisely the perturbative
    Hamiltonian constraint in vacuum. The Zerilli equation is satisfied 
    if and only if the perturbative Hamiltonian constraint is satisfied too.
    See for example Ref.~\cite{MartinGarcia:2000ze} for details.}.
  This reasoning suggests that the {\it junk} may be the macroscopic 
  manifestation of the inaccuracy in the initial-data setup at $t=0$
  (\ie of solving the linearized Hamiltonian constraint first and then interpolating),
  possibly further amplified by the wave-extraction procedure.
  This statement in itself looks confusing, because we have learned, from
  the analysis of $\psi_4$, that the Einstein (and matter) equations are
  accurately solved and that the errors made around $t=0$ due to the 
  violation of Hamiltonian constraint are relatively negligible.
  The relevant question is then: is it possible that small
  numerical errors, almost negligible in  
  $r\psi_4^{20}$, may be amplified in $\Psi^{(\rm e)}_{20}$ 
  at such a big level to produce totally nonsensical results?
  The following discussion proposes some heuristic explanation.
  
  To clarify the setup of our reasoning, let us first remind the reader 
  of the basic elements of the Abrahams-Price metric wave-extraction 
  procedure and, in particular, the role of Eq.~\eqref{eq:zerilli}.
  At a certain evolution time $t$, the numerical metric $g_{\mu\nu}(t)$ 
  is known at a certain finite accuracy on the Cartesian grid. 
  One selects coordinate extraction spheres of coordinate 
  radius $\bar{r}=\sqrt{x^2+y^2+z^2}$ on which the metric is interpolated
  via a second-order Lagrangian interpolation.
  Isotropic-coordinate systems $(\bar{r},\theta,\phi)$ naturally live on
  these spheres and thus one defines spherical harmonics.
  Then, the metric $g_{\mu\nu}$ is formally decomposed in a Schwarzschild
  ``background'' $g^0_{\mu\nu}$ plus a perturbation $h_{\mu\nu}$. 
  The next step is to choose a coordinate system in which the background
  metric is expressed. The standard approach is to use Schwarzschild 
  coordinates, although this choice actually introduces systematic errors
  that may relevantly affect the waveforms. This has been recently
  demonstrated in Ref.~\cite{Pazos:2006kz}. Although we are aware of 
  this fact, we prefer to neglect this source of error, on which we
  will further comment below. Choosing Schwarzschild coordinates means
  that one needs to compute a Schwarzschild radius $r$. This is given by
  the areal radius of the extraction two-spheres.
  Proceeding further, $h_{\mu\nu}$ is decomposed into seven 
  (gauge-dependent) even-parity 
  $(H_0,H_1,H_2,h_0^{(\rm e)},h_1^{(\rm e)},G,K)$ and 
  three (gauge-dependent) odd-parity multipoles (that we do not consider here).
  From combinations of the seven even-parity multipoles
  and of their radial derivatives, see Eqs.~(41) and (42) 
  of Ref.~\cite{Nagar:2005ea}, one obtains the 
  gauge-invariant functions $k_{\lm}$ and $\chi_{\lm}$, as well as the derivative $\de_r k_{\lm}$.
  The last step is the computation of the Zerilli-Moncrief function 
  via Eq.~\eqref{eq:zerilli}.
  Various sources of errors are present. In particular, we mention the errors originating from: 
  (i) the discretization of $g_{\mu\nu}$ (and its
  derivatives), from the numerical solution of Einstein's equations; 
  (ii) the interpolation from the Cartesian grid to the isotropic 
  grid; (iii) the computation of the metric multipoles via numerical 
  integration over coordinate (gauge-dependent) two-spheres.  
  Our aim is to investigate 
  how these inaccuracies on $(\chi_{\lm},k_{\lm},\de_r k_{\lm})$ can
  show up in $\Psi^{(\rm e)}_{\lm}$ at large extraction radii. 
  In the limit $r\gg M$, Eq.~\eqref{eq:zerilli} reads
  \begin{equation}
   \label{Z_asymptotic}
  \Psi^{(\rm e)}_{\lm} = \dfrac{2 r}{\Lambda(\Lambda -2)}\left(\chi_{\lm} -
  r\de_r k_{\lm} + \dfrac{\Lambda}{2} k_{\lm}\right),
  \end{equation}
  that is 
  \begin{equation}
  \label{zeta}
  \Psi^{(\rm e)}_{\lm} \propto r Z_{\lm}\, ,
  \end{equation}
  where $Z_{\lm}=\chi_{\lm} - r\de_r k_{\lm} + \Lambda/2 k_{\lm}$
  and $r$ is the {\it areal} radius of the coordinate two-spheres.
  The Abrahams-Price wave-extraction procedure introduces 
  then errors both on $r$ and $Z_{\lm}$. In particular, the errors
  on the (gauge-invariant) multipoles $(\chi_{\lm},k_{\lm},\de_r k_{\lm})$
  conspire in a global error on $Z_{\lm}$. In a numerical simulation one has
  $Z_{\lm}=Z_{\lm}^{\rm Exact} + \delta Z_{\lm}$ and $r = r^{\rm Schw} + \delta
  r$. Here $Z_{\lm}^{\rm Exact}$ is computed from $(\chi_{\lm}^{\rm
  Exact},k_{\lm}^{\rm Exact})$,
  that are solutions of the perturbation equation on a Schwarzschild background,
  and $r^{\rm Schw}$ is the radial Schwarzschild coordinates; 
  $\delta Z_{\lm}$ encompasses all possible errors due to the multipolar
  decomposition procedure, and $\delta r$ various inaccuracies related to the
  determination of the areal radius (\eg, those related to gauge effects).
  As a result, for the ``extracted'' Zerilli-Moncrief function we can write
  \begin{equation}
   \label{error}
   \Psi^{(\rm e)}_{\lm} \approx \Psi^{\rm Exact}_{\lm} + r^{\rm Schw}\delta Z_{\lm} + \delta
   r Z_{\lm}^{\rm Exact}.
  \end{equation}
  This equation shows that, if $\delta Z_{\lm}$ is  not zero at a certain time
  (and does not decrease in time like $1/r^{\rm Schw}$) there is a contribution 
  to the global error on $\Psi^{(\rm e)}_{\lm}$ that {\it grows} linearly 
  with the extraction radius.
  This qualitative picture is consistent with what we observe in the 3D
  waveforms: a small error on $\delta Z_{20}$ introduced at $t=0$, because
  of the approximate solution of the constraints (as indicated by 
  the analysis of $r\psi^{20}_4$ curvature waveforms), can 
  show up as a burst of radiation whose amplitude increases linearly with
  the observer location. Note that what really counts here is the error
  budget at the level of $(\chi_{20},k_{20},\de_r k_{20})$ and the related
  violation of the perturbative Hamiltonian constraint, Eq.~\eqref{iso_constraint}.
  Indeed, it might occur that, even if the 3-metric $\gamma_{ij}$ is very 
  accurate and the constraints are well satisfied at this level,
  the extraction procedures {\it adds other errors} (for example due to the 
  multipolar decomposition, computation of derivatives etc.) that may be
  eventually dominating in $\delta Z_{20}$. 
  This observation may partially justify why $r\psi_4^{20}$ is well 
  behaved, while $\Psi^{(\rm e)}_{20}$ is not.
  Finally, we note that in our evolution $\delta r$ is typically very small,
  so that we have $r^{\rm Schw}\approx r$ with good accuracy.

  Because of the complexity of the 3D wave-extraction algorithm,
  we were able neither to push forward our level of understanding, nor to 
  precisely diagnose the cause of the aforementioned 
  errors \footnote{For example, we mention, in passing, that we have 
  also tried 4th-order Lagrangian interpolation, without any visible 
  improvement on the waveform.}.
  This is now beyond the scope of the present work and will 
  deserve more attention in the future.
  By contrast, we can exploit the simpler computational framework
  offered by the 1D {\tt PerBACCo} code to ``tune'' the error $\delta Z_{\lm}$ 
  in order to produce some initial ``spurious'' 
  burst of radiation, and then possibly observe that its amplitude
  grows linearly with $\bar{r}$. In the 1D code $\delta r$ is zero 
  by construction, so that all errors are concentrated on $\delta Z_{\lm}$.
  The constrained scheme adopted in the perturbative code 
  (which is second-order convergent) allows to accurately
  compute the multipoles $(\chi_{\lm},k_{\lm})$ at every time step,
  and the Hamiltonian constraint is satisfied by construction.
  Then, $\de_rk_{\lm}$ is obtained via direct numerical differentiation 
  of $k_{\lm}$. Consequently, the error $\delta Z_{lm}$ depends 
  on the resolution $\Delta r$ as well as on the order of the
  finite-differencing representation of $\de_r k_{\lm}$. 

\begin{figure}[t]
    \begin{center}
      \includegraphics[width=85 mm]{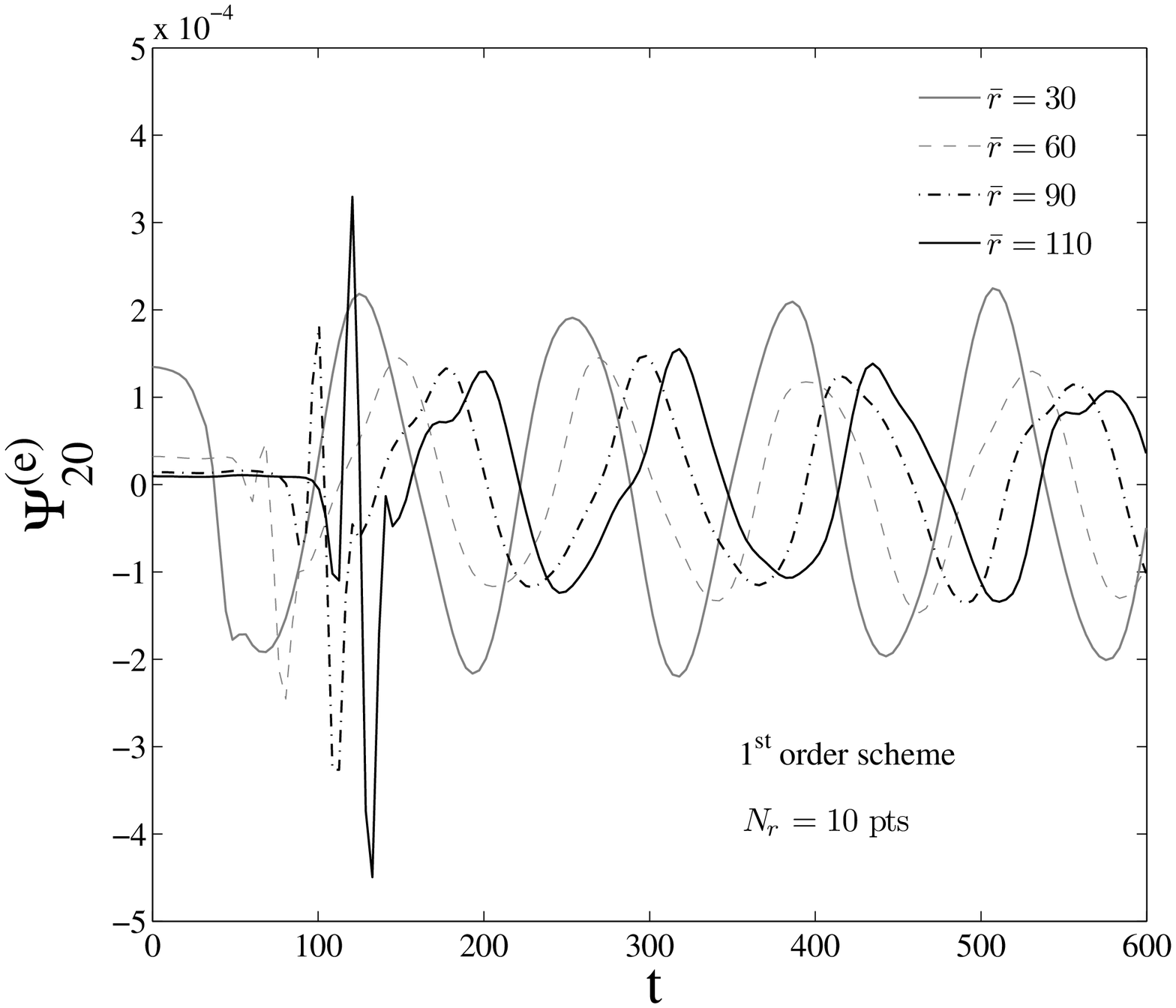} \\[1mm]
      \includegraphics[width=85 mm]{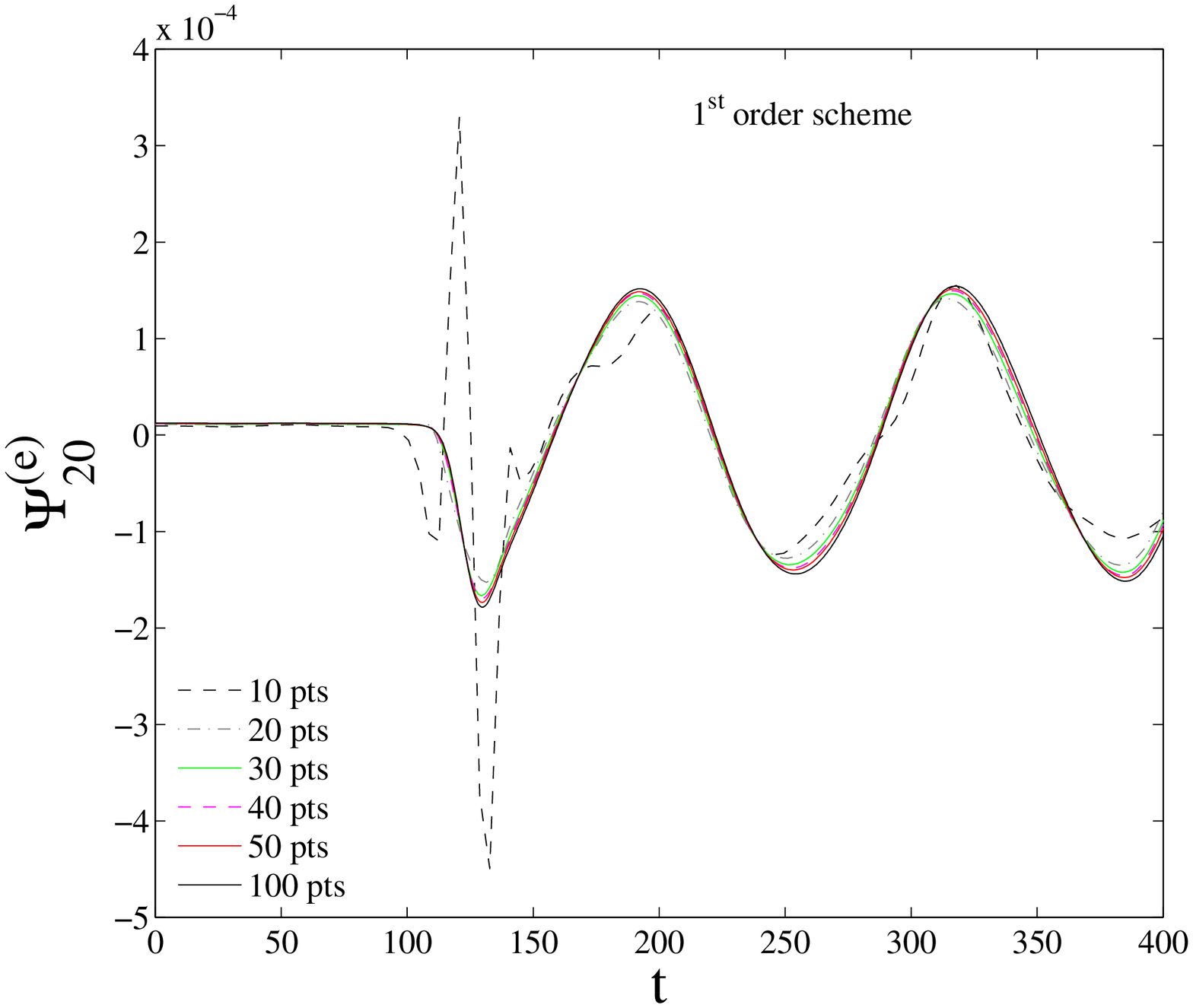}
      \caption{\label{fig:1Djunk}(color online) Metric waveforms from linear 1D evolutions.  {\bf Top panel}: 
	Low resolution simulation. A burst of {\it junk} radiation at early times 
	is present and its amplitude grows linearly with the extraction
	radius. {\bf Bottom panel}: By increasing the resolution, the initial
	{\it junk} disappears.}
    \end{center}
  \end{figure}

\begin{figure}[t]
    \begin{center}
      \includegraphics[width=95 mm]{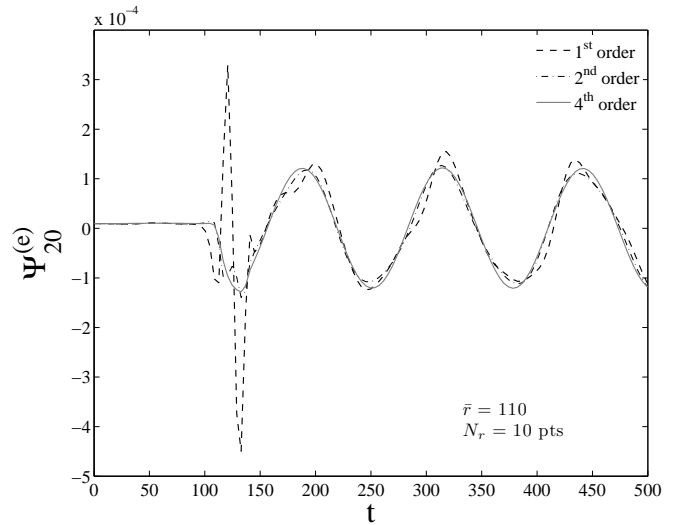}
      \caption{\label{fig:1Djunk_vs_derivative} The amount of
        {\it junk} radiation present in the waveforms extracted in the 1D code is smaller when
        higher-order differential operators are used in Eq.~\eqref{eq:zerilli} to compute the 
        the Zerilli-Moncrief function.}
    \end{center}
  \end{figure}

  In the following we shall analyze separately the effect of 
  resolution and of the approximation scheme adopted for the numerical derivatives.
  First, we approximate $\de_r k_{\lm}$ with its standard 
  first-order finite-differencing representation, 
  \ie $\de_r k_{\lm}\approx (k_{j+1}^{\lm}-k_j^{\lm})/\Delta r$ 
  and we study the behavior of the extracted
  $\Psi^{(\rm e)}_{\lm}$, computed using Eq.~\eqref{eq:zerilli},  
  versus extraction radius and resolution.
  Second, we use a fixed $\Delta r$, but  we vary the 
  accuracy of the finite-differencing representation of 
  $\de_r k_{\lm}$, contrasting first-order, second-order 
  and fourth-order stencils. The results of these two 
  analyses, for $\l=2$, $m=0$, are shown in
  Figs.~\ref{fig:1Djunk} and~\ref{fig:1Djunk_vs_derivative} respectively.
  In the top panel of Fig.~\ref{fig:1Djunk}
  $\de_r k_{20}$ is approximated at first-order, with a 
  resolution of 10 points inside the star ($\Delta r\sim 0.9$).
  This resolution 
  approximately corresponds to the resolution of the coarsest refinement
  level used in the 3D code. The extracted waveform $\Psi^{(\rm e)}_{20}$ 
  is shown at different observers, $\bar{r}\in \{30,60,90,110\}$: 
  An initial burst of {\it junk} radiation develops at early times
  and its amplitude  grows {\it linearly} with the extraction 
  radius (and it keeps growing for  $\bar{r}> 110$).
  This behavior looks identical to that found in the 3D simulations. 
  In the bottom panel of Fig.~\ref{fig:1Djunk} we focus on $\bar{r}=110$
  only, but vary the number of radial points $N_r$ inside 
  the star, namely $N_r\in\{10,20,30,40,50,100\}$. The figure shows that
  the initial {\it junk} is not present at higher resolutions
  ($N_r \geq 20$) and that the waveform converges to the exact 
  profile\footnote{As discussed in Ref.~\cite{Bernuzzi:2008rq}
  we cross-checked things also by matching the Zerilli-Moncrief function at the
  surface and evolving it with the Zerilli equation outwards. We found good
  agreement between the ``matched'' and the ``computed'' waveforms.}.
  But varying the resolution only {\it shifts} the occurrence of the burst at farther radii:
  observers at $\bar{r}\gg 110$ still see this burst appear and grow linearly with $\bar{r}$.
 
  A complementary analysis is shown in Fig.~\ref{fig:1Djunk_vs_derivative},
  where we fix the resolution at $N_r=10$ (for $\bar{r}=110$), but we 
  change the accuracy of the numerical derivative $\de_{r}k_{20}$.
  As expected, the initial {\it junk} disappears when the accuracy of the numerical 
  differential operator is increased: a second-order operator produces 
  only a small amplitude bump, that is not present when the fourth-order 
  operator is employed. 
  At this stage, the conclusion is clear: the convergence of the
  Zerilli-Moncrief function computed from the separate knowledge of the
  multipoles $k_{\lm}$ and $\chi_{\lm}$ is a delicate issue that must
  be analyzed with care according to the physical problem under consideration.
  The violation of the perturbative Hamiltonian constraint and, in particular, the
  accuracy of the numerical derivative $\de_r k_{\lm}$ 
  (note that we refer to the induced violation at the level of the 
  wave extraction and {\it not} at that of the solution of the perturbation
  equations) seems to play an important role in the convergence properties of the waveforms.
  The main conclusions of the aforementioned numerical tests are: (i) The errors in 
  $\Psi^{(\rm e)}_{20}$ seem to behave like suggested in 
  Eq.~\eqref{error}; (ii) The phenomenon occurs in the same way
  in both the 1D and 3D code, although the fine details of the oscillation
  are different.

  Focusing on the 1D {\tt PerBACCo} code, an accurate 
  $\Psi^{(\rm e)}_{\lm}$ is obtained using sufficiently high resolution
  ($N_r= 300$) as well as a fourth-order representation for $\de_r
  k_{\lm}$. These prescriptions are accurate enough for the problem addressed
  in this work, although they may not be sufficient for other stellar models or
  other initial perturbations. For example, using the {\tt PerBACCo} code, with the
  same initial data setup discussed here, in order to study
  the time evolution of perturbations of stars with realistic 
  EOS proved that higher resolutions are typically needed to produce convergent waveforms of comparable 
  accuracy~\cite{Bernuzzi:2008fu}. 
  Likewise, for a polytropic EOS and initial data given by a Gaussian 
  pulse in $\Psi^{(\rm e)}_{20}$, the same Ref.~\cite{Bernuzzi:2008fu} 
  showed that {\it at least} fourth-order accuracy in $\de_r k_{20}$ is needed 
  in order to have a consistent extraction of $\Psi^{({\rm e})}_{20}$ {\it already} 
  at $t=0$ (see Appendix~A of Ref.~\cite{Bernuzzi:2008fu}).
  This suggests that the presence of linearly
  growing {\it junk} radiation in the computation of $\Psi^{(\rm e)}_{\lm}$ 
  from the multipoles $(k_{\lm},\chi_{\lm})$ can appear ubiquitously
  in the time evolutions of the perturbation equations with
  the {\tt PerBACCo} code. The presence of this {\it junk} 
  in $\Psi^{(\rm e)}_{\lm}$ is the macroscopic manifestation of the 
  violation of the perturbative Hamiltonian constraint due to errors
  (notably, in the discretization of the derivatives) introduced in the wave-extraction
  procedure. These (typically small) numerical errors are eventually 
  magnified by the presence of an overall  $r$ factor in
  Eq.~\eqref{Z_asymptotic}. Note that this phenomenon occurs even if
  the computation of the multipoles $(k_{\lm},\chi_{\lm})$ is very
  accurate and the Hamiltonian constraint is satisfied by construction
  in the evolution algorithm.
  The analysis that we have presented here suggests that either increasing 
  the resolution or, more reasonably, implementing higher-order differential 
  operators in the perturbative ``extraction'' procedure are viable proposals 
  to compute convergent waveforms.
  \begin{figure}[t]
    \begin{center}
      \includegraphics[width=85 mm]{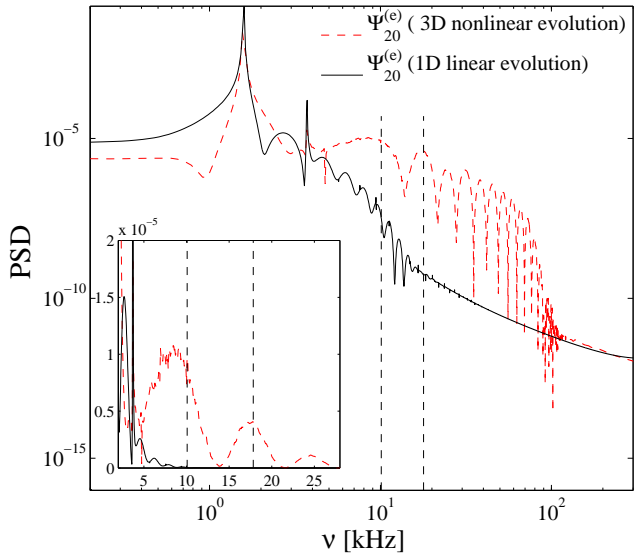}
      \caption{\label{fig:PSD_junk} (color online) PSD of the 3D
      metric waveforms (dashed line) extracted \'a la Abrahams-Price (at
      $\bar{r}=110$) and the corresponding
      1D waveform (solid line) for a simulation with perturbation $\lambda=\lambda0$. 
      The Fourier spectrum of the {\it junk} radiation is compatible with 
      some $w$-mode frequencies.
      }
    \end{center}
  \end{figure}

  In the 3D case the situation is more involved and we
  have not succeeded in making statements as solid as in the 1D case. 
  We can only rely on analogies: 
  (i) The appearance of the {\it junk} occurs in a way similar to the 1D
  case when the accuracy of the 1D Zerilli function is low; 
  (ii) The two time evolutions look qualitatively very similar. 
  Yet, it is not technically possible to use in the 3D code resolutions
  equivalent to those of the 1D code. 
  By analogy with our perturbative results,
  we can only conclude that it is not unreasonable that 
  the {\it junk} in the 3D waveforms is the macroscopic manifestation of 
  inaccuracies hidden in the implementation of the Abrahams-Price 
  wave-extraction procedure. The analysis presented here points out
  that such metric wave-extraction procedures require typically more subtle care
  than expected and these subtleties must be kept in mind in developing
  more modern wave-extraction routines.

  The PSD of $\Psi^{(\rm e)}_{20}$ in both the 1D case (solid line) and 3D 
  case (dashed line) is displayed in Fig.~\ref{fig:PSD_junk}. 
  The perturbation is $\lambda=\lambda0$ and the extraction radius is $\bar{r}=110$.
  The spectrum of the 3D Zerilli waveform is consistent 
  with what we observed in $r\psi_4^{20}$ below $10$~kHz, 
  but it looks different at higher frequencies 
  (compare it with that of the ``integrated'' $\Psi^{(\rm e)}_{20}$
  in Fig.~\ref{fig:PSD_psi4}): here the PSD shows broad peaks
  attributable to the initial part of the waveform.
  Recovering the reasoning started in the previous section about the presence of the
  $w$ modes, we can observe that the frequencies contained in the {\it junk}
  are compatible with the $w$-mode frequencies $\nu_{w_1}$ and $\nu_{w_2}$
  (indicated by dashed vertical lines in Fig.~\ref{fig:PSD_junk}).
  However, since such frequencies belong to an unphysical 
  part of the waveform, we prefer to consider them unphysical as well.

  \begin{figure}[t]
    \begin{center}
      \includegraphics[width=85 mm]{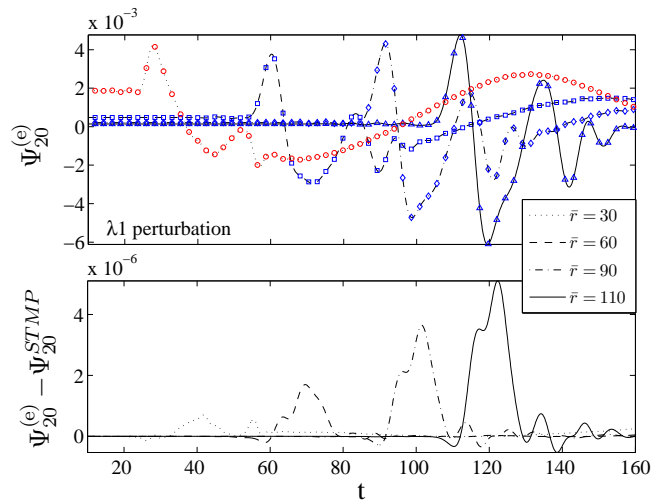}
      \caption{\label{fig:weg} (color online) The {\bf top panel} compares 
       the standard Zerilli-Moncrief $\Psi^{(\rm e)}_{20}$ (depicted with
       lines) and the generalized
       $\Psi^{\rm STMP}_{20}$ (depicted with point markers) for the first part 
       of the gravitational-wave signal in a simulation with 
       perturbation $\lambda=\lambda1$. The waves extracted at four radii
       $\bar{r}\in\{30,60,90,110\}$ are shown.
       The {\bf bottom panel} shows that the present  
       {\it junk} radiation negligibly depends on 
       the choice of the coordinates of the background metric.}
    \end{center}
  \end{figure}

  We conclude by mentioning, in passing, that the initial {\it junk} 
  radiation is {\it essentially not related} to the systematic error 
  introduced by fixing Schwarzschild coordinates for the background 
  metric $g^{0}_{\mu\nu}$. This fact is suggested by Fig.~\ref{fig:weg},
  where we contrast the standard Zerilli-Moncrief function $\Psi^{(\rm
  e)}_{20}$ (which assumes Schwarzschild coordinates for the background)
  with the {\it generalized} $\Psi^{\rm STMP}_{20}$ one based on the 
  Sarbach-Tiglio~\cite{Sarbach:2001qq} and Martel-Poisson~\cite{Martel:2005ir} 
  perturbation formalism, which does not require any gauge-fixing condition
  for the background submanifold $M^2$.
  This particular simulation was performed over a grid with three 
  refinement levels and cubic boxes with limits [-120, 120], [-24,24] and
  [-12,12]. The resolution of each box is coarser than in
  the previous simulations, namely $\Delta_{xyz}=1.875$, $0.9375$ and $0.46875$
  respectively.
  Evidently, with this resolution the waveforms are less accurate,
  but we do not mind at this stage, since we are interested 
  in an intrinsic comparison between extraction procedures at fixed resolution.
  The function $\Psi^{\rm STMP}_{\lm}$ that we use is given by the   
  straightforward\footnote{With this we mean that we do not take into account
  any time dependence of the background metric due to
  coordinate effects. This possibility can be anyway easily taken into account 
  by the formalism. We postpone to a future work the related discussion~\cite{bernuzzi_prep}.} 
  implementation of Eq.~(4.23) of Ref.~\cite{Martel:2005ir}. Note that 
  this expression is equivalent to the combination of 
  Eqs.~(20), (25), (26) and (27) of Ref.~\cite{Sarbach:2001qq}.
  The top panel of Fig~\ref{fig:weg} displays $\Psi^{(\rm e)}_{20}$ (lines)
  and $\Psi^{\rm STMP}_{20}$ (point markers) 
  for observers at $\bar{r}\in\{30,60,90,110\}$. It highlights that 
  the differences in the early-time part of the waveforms are very small.
  By contrast, the bottom panel of the figure, showing the 
  difference $\Psi^{(\rm e)}_{20}-\Psi^{\rm STMP}_{20}$, indicates 
  that removing (part of) the systematic errors generates some improvement,
  but this is too small to be of any relevance.
  This analysis suggests that the inaccuracies in the early-time part
  of the waveform are essentially not related to the specific computation
  of the (generalized) Zerilli function, but rather connected 
  to the underlying multipolar extraction infrastructure 
  (grid setup, approximate solution of the constraints, 
  interpolation procedures, computation of the derivatives of 
  the metric etc.), on which we have relatively little control.
  A comprehensive analysis of the problems related to the 
  generalized extraction procedure will be presented elsewhere~\cite{bernuzzi_prep}.
  We remark, however, that systematic effects that are very small in
  our physical system, as emphasized by our Fig.~\ref{fig:weg}, 
  may be not small in other situations, 
  as found in Ref.~\cite{Pazos:2006kz}. For this reason, we emphasize
  that the formalism of Refs.~\cite{Sarbach:2001qq,Martel:2005ir} is
  the actual {\it correct metric formalism} to extract waveforms
  out of a numerical space-time that can be considered a small deformation of 
  the Schwarzschild one. As such, it must be taken into account
  properly in numerical codes.
  
  \subsection{Generalized quadrupole-type formulas}
  \label{sbsc:WaveQuad}
  \begin{figure*}[t]
    \begin{center}
      \includegraphics[width=90 mm]{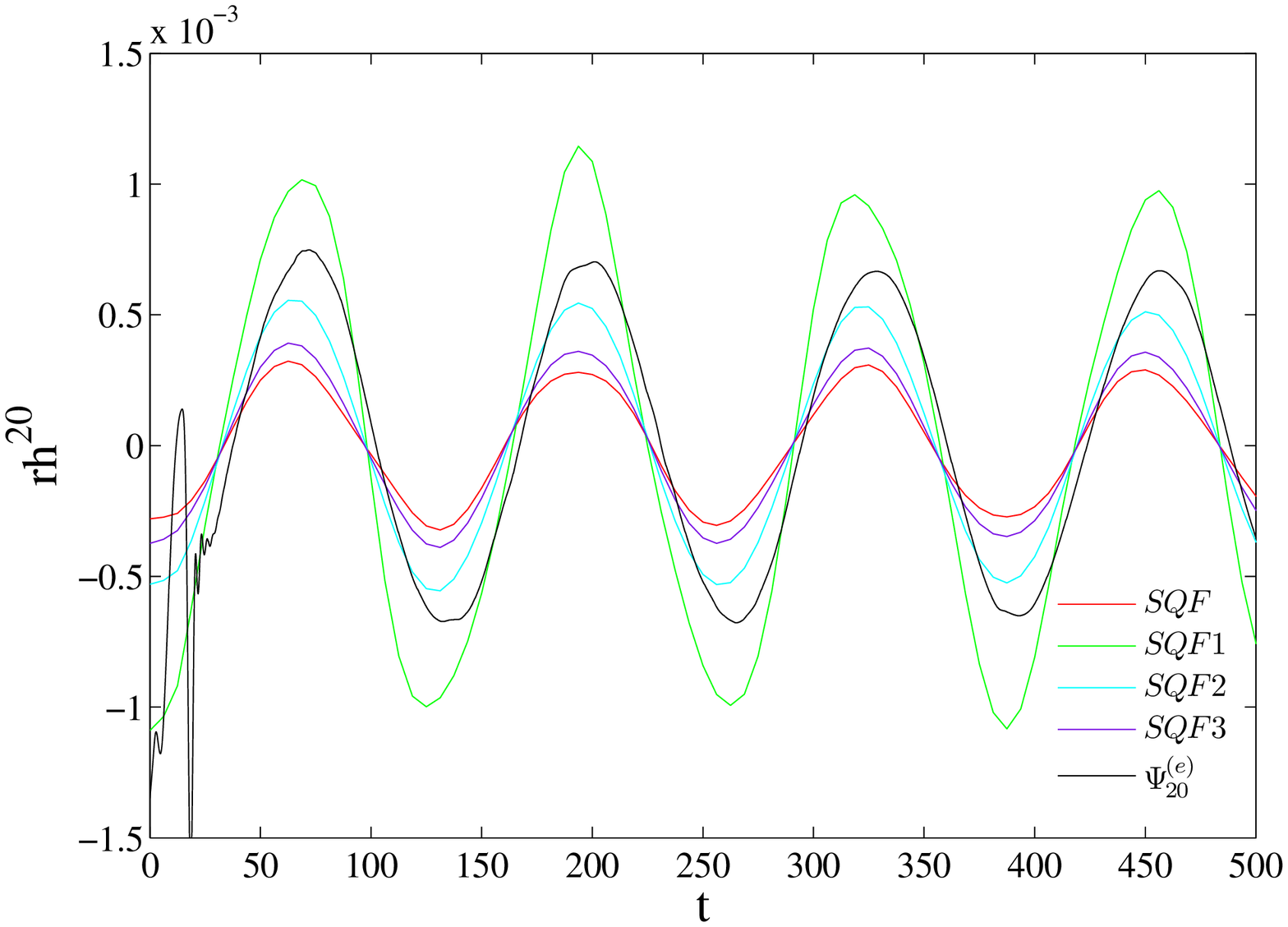}
      \includegraphics[width=80 mm]{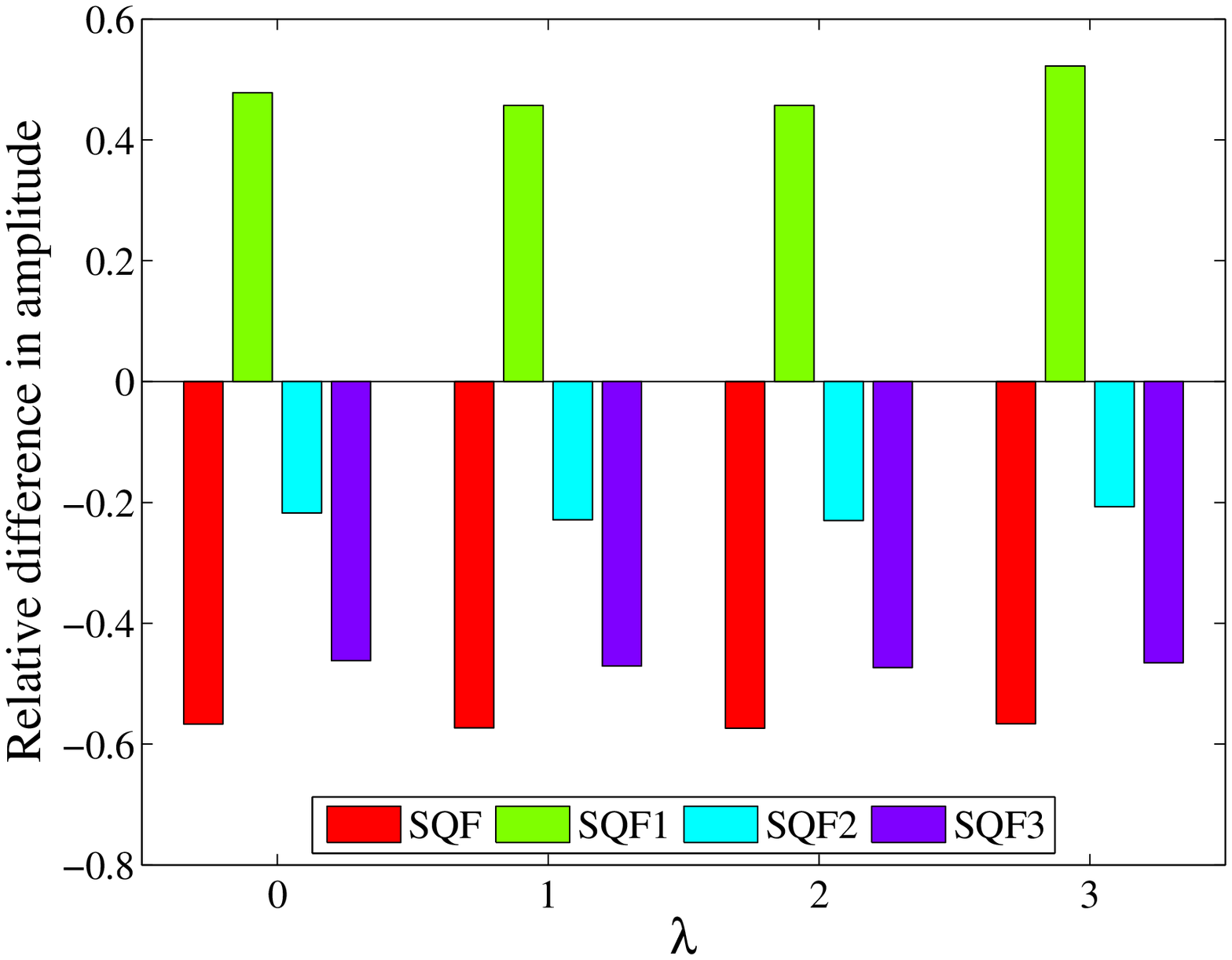}
      \caption{\label{fig:h20:quadrupole} (color online) Comparison between different wave-extraction procedures in
        the 3D simulations, in particular between the SQFs and the gauge-invariant wave-extraction
        procedure. The {\bf left panel} shows the waveforms. The frequencies almost coincide, while the amplitudes
        are strongly underestimated in all cases except SQF1, where the
        amplitude is strongly overestimated. The {\bf right panel} shows the relative difference between the
        amplitudes of the various SQFs and the Abrahams-Price wave extraction. These amplitudes are
        estimated with the fit procedures.}
    \end{center}
  \end{figure*}

  Finally, we study the performances of the various generalized 
  quadrupole-type formulas that we have introduced in Sec.~\ref{sec:we}.
  The results of our analysis are shown in Fig.~\ref{fig:h20:quadrupole}.
  The left panel of the figure displays $rh^{20}$ 
  waveforms obtained via the 
  SQF1, SQF2, SQF3 and SQF4 [see Eqs.~(\ref{sqf}--\ref{sqf3})]
  for perturbation $\lambda=\lambda0$.
  The right panel complements this information by showing
  (for several perturbation magnitudes $\lambda$)
  the relative difference in amplitude between the various SQFs and the corresponding 
  gauge-invariant Zerilli-Moncrief function $\Psi^{(\rm e)}_{20}$.
  This analysis highlights that the quadrupole formula gives an 
  excellent approximation to the phasing of the actual signals. 
  By contrast, there is a systematic over- or under-estimation of the 
  amplitude depending on the choice of SQF.
  
  A related observation is that the discrepancy between the quadrupole formula
  and the gauge-invariant waveform is not due to 
  the fact that waveforms are extracted at a finite radius. 
  Our results are consistent with those of Shibata and 
  Sekiguchi~\cite{Shibata:2003aw}, who performed an analysis 
  similar to ours (and also considered uniformly rotating stars), 
  but without the possibility of contrasting their results
  with linear evolutions. In this respect, the main conclusion of 
  Ref.~\cite{Shibata:2003aw} was that, although the amplitude of 
  the waveform is systematically underestimated 
  by the quadrupole formula~(\ref{eq:multipoles_20}), it is however 
  sufficiently accurate to capture both the frequency and the phasing 
  (that are the most important quantities for detection) of the waveforms 
  in a proper way. These results are fully confirmed here using totally
  different codes.

  \subsection{Nonlinearities}
  \label{sbsc:nnlin}
    
  In this section we comment on the onset of nonlinear effects for high values of the 
  initial perturbation amplitude $\lambda$, showing, 
  for the first time using full GR simulations, 
  evidences for mode couplings and for the appearance of nonlinear harmonics. 

  \begin{figure}[t]
      \begin{center}
      \includegraphics[width=85 mm]{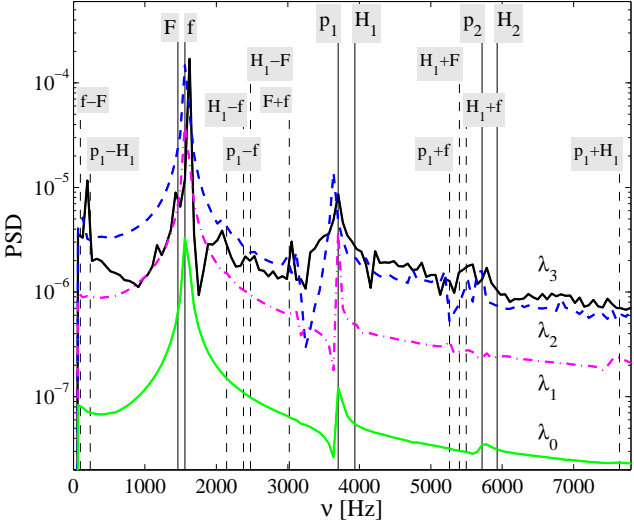}
      \end{center}
      \vspace{-8mm}
      \caption{\label{fig:couplings} (color online) PSD of the quantity
	$\langle\rho\rangle_{20}(t)$ [see Eq.~\ref{eq:rholm}]
        for different values of the initial perturbation amplitude $\lambda$. 
        The spectra of the ($\l=2$, $m=0$) mode obtained in simulations with 
	larger perturbations contain more frequencies, which originate from
	the nonlinear couplings with the overtones and with the radial modes.
      }    
  \end{figure}
  
  In the linear regime ($\lambda=\lambda0$) the star is oscillating at, essentially, the frequency
  of the fundamental quadrupolar proper fluid (quasinormal) mode of pulsation.  The principal
  linear modes excited are thus the $(\l,m)=(2,0)$ one and its overtones.  For growing values of the
  initial perturbation, we observe that in 3D simulations, differently from the linear ones, the
  amplitude of the multipole $\Psi^{(\rm e)}_{20}$ does not increase proportionally to $\lambda$
  but, instead, is progressively reduced (see Fig.~\ref{fig:resAmp}). This fact could be interpreted as
  the results of a typical phenomenon in nonlinear systems in which linear modes \emph{couple},
  generating nonlinear harmonics.  Naively, one could think that the ``energy'' associated to the
  $\l=2$ mode is redistributed to the others while the system departs from the linear
  regime\footnote{We would like to stress that this picture is not so simple and rigorous, in
    particular there is no proof of the completeness of the star quasinormal modes (even in the
    nonrotating case) and the definition of an energy per mode is definitely not straightforward.
    See the discussion in Ref.~\cite{1999CQGra..16R.159N}.}.  As we saw in Sec.~\ref{sbsec:radial},
  radial modes of oscillation are already present in the evolution of the equilibrium models.  As a
  consequence, we expect to reveal couplings between nonradial and radial modes ($F$ and its
  overtones $H_1,H_2,...$) as a result of the onset of some nonlinear effect.  In addition, we detect
  signals in multipoles of $\Psi^{(\rm e)}_\lm$ with $\ell=4,6$ and $m=0,4$, that are even-parity
  axisymmetric modes and nonaxisymmetric modes triggered by the Cartesian grid.  The amplitudes are
  very weak compared to those of the $f$ mode for any value of $\lambda$, typically 2 orders of magnitude smaller for
  $\l=4,m=0$ and 3 orders of magnitude for the others, but in
  principle they are present and must be considered.  As far as the odd-parity modes with
  $m=1,2,3$, and $\l=3,5$ are concerned, they are all forbidden by the symmetry imposed on the computational
  domain (octant).

  As a strategy to study nonlinearities, we consider the rest-mass--density projections:
  \begin{equation}
    \label{eq:rholm}
    \langle\rho\rangle_{\lm} (t) \equiv \int d^3x \rho(t,\mathbf{x})Y^*_{\lm}
  \end{equation}
  and we apply to them the Fourier analysis. Like all the global variables, $\rho$ contains
  all the frequencies of the system. Its projections in Eq.~\eqref{eq:rholm} allow to separate  
  the contribution of each mode $(\l,m)$.
  Fig.~\ref{fig:couplings} shows the power spectrum of
  $\langle\rho\rangle_{20}$ for the four different
  values of $\lambda$. 
  The signal for $\lambda=\lambda0$ contains the 3
  frequencies of the linear modes $f$, $p_1$ and $p_2$. 
  The same happens for $\lambda=\lambda1$ and $\lambda=\lambda2$: the amplitudes 
  of the linear modes grow linearly with $\lambda$ and some new frequencies
  are present with small power for $\lambda=\lambda2$. In the case of $\lambda=\lambda3$, 
  which corresponds to a pressure perturbation of 10\% of the central TOV
  value, the spectra is rich of nonlinear harmonics. Most of them 
  can be recognized as due to \emph{weak couplings}, \ie sums and differences 
  of linear mode frequencies $\nu_1\pm\nu_2$, also called \emph{combination
  tones}. 
  In particular we identify the nonlinear harmonics of the 
  $f$ mode and its overtone $f\pm p_1$ and many
  frequencies $f\pm F$, $f\pm H_1$ and $p_1\pm H_1$ due to the 
  radial and nonradial mode couplings.
  Such couplings have been previously and extensively studied in 
  Refs.~\cite{Font:2000rd,Stergioulas:2003ep,Dimmelmeier:2005zk} using 
  the Cowling approximation as well as the conformally flat  approximation to GR.
  In addition, the couplings between radial and nonradial modes have been
  studied in detail in  Ref.~\cite{Passamonti:2007tm} by means of a
  second-order perturbative approach. Note how our fully general-relativistic
  results are consistent with all these studies.

  The projection $\langle\rho\rangle_{00}$ describes essentially the radial
  mode of pulsations; analyzing this quantity instead of
  $\langle\rho\rangle_{20}$ gives analogous results in term of couplings.
  From the analysis of higher multipoles we compute the frequencies of the
  linear modes, finding $\nu=2404$ Hz for $\l=4$ 
  and $\nu=2988$ Hz for $\l=6$. No couplings can be clearly recognized in these data. 
  We stress that frequencies of the nonaxisymmetric modes ($m=4$) are the same as the axisymmetric ones 
  because the star is nonrotating and modes are degenerate in $m$.

  \section{Conclusions}
  \label{sec:concl}
  
  We have compared various gravitational-wave--extraction methods that are
  nowadays very popular in numerical-relativity simulations: (i) the
  Abrahams-Price~\cite{Abrahams:1995gn} technique based on the 
  gauge-invariant Regge-Wheeler-Zerilli-Moncrief perturbation theory
  of a Schwarzschild space-time; (ii) the extraction method based on
  Weyl curvature scalars, notably the $\psi_4$ function; (iii) 
  some (variations of) quadrupole-type formulas.
  We have applied these methods to extract gravitational radiation from
  3D numerical-relativity simulations of the very controlled system 
  represented by a neutron star (with polytropic EOS), that is oscillating 
  nonradially due to an initial pressure perturbation. The simulations 
  have been performed via the {\tt Cactus-CCATIE-Carpet-Whisky} general-relativistic 
  nonlinear code. This code evolves the full set of Einstein
  equations in full generality in the three spatial dimensions.
  The accuracy of the waveforms extracted from the simulations, using 
  the three methods recalled above, has been assessed (for small
  perturbations) via a comparison with waveforms (assumed to be exact) 
  computed by means of the {\tt PerBACCo} perturbative code. This
  code is designed to evolve, in the time domain, the Einstein equations
  linearized around a TOV background. It is 1+1-dimensional (\ie one temporal 
  and one spatial dimension) and adopts a constrained evolution scheme. 
  This latter choice allows for the computation of very long and very 
  accurate time series and similarly accurate waveforms. 
  
  The initial pressure perturbation $\delta p$ is given as an ``approximate'' 
  eigenfunction of the star, whose maximum is a fraction of the central
  TOV pressure $p_c$. We focused only on $\l=2$, $m=0$, quadrupolar deformations,
  but we analyzed four values of the perturbation in order to cover the transition from 
  the linear to the nonlinear oscillatory regimes.
  We have first presented results of simulations done using only the 1D {\tt
  PerBACCo} code to assess the accuracy of our exact waveforms. 
  We have performed very long (about 1~s) and accurate simulations to
  extract both mode frequencies and damping times. We have analyzed finite-radius 
  effects, finding that observers should be placed at extraction 
  radius $r>200M$ in order to have amplitude errors below 1.6\%.

  In doing 3D simulations in the perturbative regime, 
  ($10^{-3}\lesssim {\rm max}(\delta p/p_c)\lesssim 10^{-2}$), we have found 
  that {\it both} metric and curvature wave-extraction techniques generate
  waveforms that are consistent, both in amplitude and phasing, 
  with the perturbative results.
  Each method, however, was found to have drawbacks. On one hand, the 
  Zerilli-Moncrief function presents an unphysical burst in the early part
  of the waveform; on the other hand, the $\psi_4$ scalar requires a polynomial correction to 
  obtain the corresponding metric multipole.
  Our conclusion is that, in our setup, one needs both extraction 
  methods to end up with accurate waveforms.

  For larger values of the initial perturbation amplitude, nonlinear effects
  in the 3D general-relativistic simulations are clearly present.
  The effective relative amplitude of the main modes of the extracted gravitational 
  wave is smaller for larger amplitudes of the initial perturbation, because of mode 
  couplings.
  The Fourier spectra of the rest-mass--density projections [see Eq.~\eqref{eq:rholm}]
  highlight that couplings between radial and quadrupolar fluid modes are
  present. Our study represents the first confirmation, in fully
  general-relativistic simulations, of the results of
  Ref.~\cite{Passamonti:2007tm}, obtained via a perturbative approach.

  In addition, we have shown that the (non-gauge-invariant) generalizations of 
  the standard Newtonian quadrupole formula that we have 
  considered can be useful tools to obtain accurate estimates of 
  the frequency of oscillation. By contrast, amplitudes are always
  significantly under/overestimated, consistently with precedent 
  observations of Refs.~\cite{Shibata:2003aw,Nagar:2005cj}.

  Finally, we discussed in detail some systematic errors that
  occur in the early part of the waveform extracted \'a la Abrahams-Price.
  These errors show up, in the early part of the Zerilli-Moncrief function, 
  in the form of a burst of {\it junk} radiation
  whose amplitude grows linearly with the extraction radius. 
  We have proposed some heuristic explanation of this fact and reproduced 
  a similar behavior in low-accuracy perturbative simulations. 
  Globally, our conclusion is that the extraction of the Zerilli-Moncrief
  function from a numerical-relativity simulation can be a delicate issue:
  small errors can conspire to give totally nonsensical results.
  Typically, these errors will show up as parts of the waveform whose 
  amplitude grows with the observer's radius.
  We have also implemented the generalized wave-extraction approach based
  on the formalism
  of Refs.~\cite{Sarbach:2001qq,Martel:2005ir,Pazos:2006kz,Korobkin:2008ji}, 
  without any evident benefit.
  Note, however, that these kind of problems encountered with the
  Abrahams-Price wave-extraction procedure (as well as with its generalized
  version) seem to appear {\it specifically} in the presence of matter.
  In binary black-hole coalescence simulations curvature and metric
  waveforms seem to be fully consistent~\cite{Pollney:2007ss}.
  This last remark leads us to suggest that the Abrahams-Price 
  wave-extraction technique, a ``standardized'' and very basic 
  procedure and infrastructure that has been developed long ago (and tested
  at the time) for specific applications to black-hole physics, should be
  rethought and reanalyzed when the Einstein equations are coupled to matter.
  For this reason, in the presence of matter, since systematic errors could 
  be hard to detect and are present already in the simplest cases, we strongly
  encourage the community to make use of {\it both} wave-extraction techniques 
  (curvature as well as metric perturbations) and to be always prepared to
  expect inaccuracies in the metric waveforms.
  In addition, concerning the many advantages related to extracting the metric waveforms
  directly from the space-time, we believe that it is also urgent and important 
  for the community to have reliable implementations of the
  Abrahams-Price technique based on the 
  Sarbach-Tiglio-Martel-Poisson~\cite{Sarbach:2001qq,Martel:2005ir,Pazos:2006kz,Korobkin:2008ji} formalism.

  
  \section*{Acknowledgments}
  
  We are grateful to N.~Stergioulas, who posed the questions that 
  eventually led to this article, and to L.~Rezzolla for discussion, 
  suggestions and constructive criticisms. 
  A.~Nagar thanks T.~Damour for fruitful discussions.
  S.~Bernuzzi, R.~De~Pietri and  A.~Nagar are 
  grateful to the Albert-Einstein-Institut
  for hospitality during the preparation of this work.
  S.~Bernuzzi and G.~Corvino thank IHES for hospitality
  during the development of this work.
  A.~Nagar and R.~De~Pietri also thank L.~Castellani and P.~Fr\'e for
  hospitality at the Dipartimento di Fisica, Universit\`a di Torino.
  The activity of A.~Nagar at IHES is funded by INFN.
  This research was  supported in part by INFN Iniziativa specifica 
  OG51; University of Parma Grant No. FIL0719037; the DFG Grant 
  No. SFB/Transregio 7; the JSPS Grant-in-Aid for Scientific Research 
  (19-07803).
  Computations have been performed on the INFN Beowulf clusters 
  {\tt Albert} at the University of Parma and on the Peyote and
  Damiana clusters at the Albert-Einstein-Institut. 
  
  
  \bibliography{StarWaves}
  
  
\end{document}